\documentclass[aps,prd,twocolumn,floatfix,nofootinbib,groupedaddress,superscriptaddress]{revtex4-1}
\pdfoutput=1 
\usepackage{amsmath, amssymb, graphics, setspace}
\newcommand{\mathsym}[1]{{}}
\newcommand{\unicode}[1]{{}}
\usepackage{textcomp}
\usepackage{eurosym}
\usepackage{amsfonts}
\usepackage{array}
\usepackage{amsthm}
\usepackage{changes}
\usepackage{lipsum}
\definechangesauthor[name={Per cusse}, color=orange]{per}
\usepackage{bm}
\usepackage{palatino}

\usepackage{hyperref}
\usepackage{graphicx}
\usepackage{amsmath, amsfonts, amssymb}
\usepackage{bm, array}
\usepackage{mdframed}
\usepackage{grffile}
\usepackage{amssymb}
\usepackage{caption}
\usepackage{ulem}
\usepackage[T1]{fontenc}

\def\be{\begin{equation}}
\def\ee{\end{equation}}
\def\beq{\begin{eqnarray}}
\def\eeq{\end{eqnarray}}

\begin{document}

\title{Falsifying cosmological models based on a non-linear electrodynamics}

\author{Ali \"{O}vg\"{u}n}
\email{ali.ovgun@pucv.cl}
\homepage[]{https://www.aovgun.com}
\affiliation{Instituto de F\'{\i}sica, Pontificia Universidad Cat\'olica de
Valpara\'{\i}so, \\ Casilla 4950, Valpara\'{\i}so, Chile}

\affiliation{Physics Department, Arts and Sciences Faculty, Eastern Mediterranean University,\\ Famagusta, North Cyprus via Mersin 10, Turkey}

\affiliation{Physics Department, California State University Fresno, Fresno, CA 93740,
USA.}
\affiliation{Stanford Institute for Theoretical Physics, Stanford University, Stanford,
CA 94305-4060, USA}

\author{Genly Leon}
\email{genly.leon@ucn.cl} 
\affiliation{Departamento de Matem\'aticas, Universidad Cat\'olica del Norte,\\ Avda. Angamos 0610, Casilla 1280 Antofagasta, Chile}

\author{Juan Maga\~na}
\email{juan.magana@uv.cl}
\affiliation{Instituto de F\'{\i}sica y Astronom\'{\i}a, Facultad de Ciencias,
Universidad de Valpara\'{\i}so, Avda. Gran Breta\~na 1111, Valpara\'{\i}so, Chile.}

\author{Kimet Jusufi}
\email{kimet.jusufi@unite.edu.mk}
\affiliation{Physics Department, State University of Tetovo,\\ Ilinden Street nn, 1200, Tetovo, 
Macedonia}
\affiliation{Institute of Physics, Faculty of Natural Sciences and Mathematics,\\ Ss. Cyril and Methodius University, Arhimedova 3, 1000 Skopje, Macedonia}


\date{\today}

\begin{abstract}
Recently, the nonlinear electrodynamics (NED) has been gaining attention to generate primordial magnetic fields in the Universe and also to resolve singularity problems. Moreover, recent works have shown the crucial role of the NED on the inflation. This paper provides a new approach  based on a new model of NED as a source of gravitation to remove the cosmic singularity at the big bang and explain the cosmic acceleration during the inflation era on the background of stochastic magnetic field. Also, we found a realization of a cyclic Universe, free of initial singularity, due to the proposed NED energy density. In addition, we explore whether a NED field without or with matter can be the origin of the late-time acceleration.
For this we obtain explicit equations for $H(z)$ and  perform a MCMC analysis to constrain the NED parameters by using $31$ observational Hubble data (OHD) obtained from cosmic chronometers covering the redshift range $0 < z < 1.97$;  and with the joint-light-analysis (JLA) SNIa compilation consisting in $740$ data points in the range $0.01<z<1.2$.
All our constraints on the current magnetic field give $B_{0}\sim 10^{-31}\mathrm{cm^{-1}}$, which are larger than the upper limit $10^{-33}\mathrm{cm^{-1}}$ by the Planck satellite implying that NED cosmologies could not be suitable to explain the Universe late-time dynamics. However, the current data is able to falsify the scenario at late times. Indeed, one is able to reconstruct the deceleration parameter $q(z)$ using the best fit values of the parameters obtained from OHD and SNIa data sets. If the matter component is not included, the data sets predict an accelerated phase in the early Universe, but a non accelerated Universe is preferred in the current epoch. When a matter component is included in the NED cosmology, the data sets predict a $q(z)$ dynamics similar to that of the standard model. Moreover, both cosmological data favor up to $2\sigma$ confidence levels an accelerating expansion in the current epoch, i.e., the Universe passes of a decelerated phase to an accelerated stage at redshift $\sim 0.6$. Therefore, although the NED cosmology with dust matter predict a value $B_{0}$ higher than the one measured by Planck satellite, it is able to drive a late-time cosmic acceleration which is consistent with our dynamical systems analysis and it is preferred by OHD and SNIa data sets. 
\end{abstract}


\date{\today}

\maketitle

\section{Introduction}

Universe started with a Big Bang, which had a singularity that all the laws of physics would have broken down \cite{Carroll,haw}. Today, the first thing that is known that the two great theories of physics such as quantum mechanics and general relativity mostly work very well, except in some extreme conditions like the Big Bang \cite{rev}. There is something which is clearly still missing to explain this singularity \cite{muk}. Recently, cosmological models using non-linear electromagnetic fields (NEF) have been gain interest to remove singularity problem of the Universe at the Big Bang and also singularities of curvature invariants \cite{kruglov1,kruglov2,kruglov3,ao,sadia}. The Standard Cosmological Model (SCM) based on Friedmann- Robertson-Walker  (FRW)  geometry, also known as $\Lambda CDM$ model, does not produce any solution for the singularity problem at the beginning of the Universe \cite{darkenergy}. The SCM has a problem of singularities. If the Maxwell equations are intelligently modified, these singularities can be resolved. There are some cosmological models known as magnetic Universe that has no singularity because of the nonlinear modification of the Maxwell electrodynamics at strong fields such as early universe \cite{magneticU1,novello0}, because of the background of the conformally flat Robertson-Walker metric, but the cost is the break up of the conformal invariance of Maxwell theory  \cite{novello0,novello1,novello2,vol1,novello3,novello4,novello5,w,w1,Campanelli:2007cg}. 

Moreover, today it is widely accepted fact among the physicist that the Universe is accelerating. The idea of an accelerating Universe is supported  and  confirmed  by type Ia supernovae and the cosmic microwave background (CMB) \cite{Carroll,darkenergy,riess,perl}.  However the reason of the acceleration of the Universe is not entirely clear, nevertheless a number of solutions have been proposed \cite{darkenergy,ao,lat,GarciaSalcedo:2002jm,cos1,cos2,born1,camara,eliz,coupling,durmus,Azri:2018qux,beck,nonm1,nonm2,horava,const,Leon:2009rc,Xu:2012jf,Leon:2012mt,Leon:2013qh,Kofinas:2014aka,Leon:2012vt,Fadragas:2013ina,Leon:2015via,Pulgar:2014cba,Leon:2014yua,Giacomini:2017yuk,Aydiner:2016mjw}.  One such solution is to introduce the cosmological constant in the Einstein's field equations.  In this scenario, the acceleration of Universe is driven by dark energy (DE)  which can be thought of as a kind of space-filling fluid with constant energy density through the Universe \cite{darkenergy,Joyce:2016vqv}. Another exotic form  of matter proposed as a DE candidate is to consider a scalar field known as the quintessence \cite{Tsujikawa:2013fta}.  On the other hand, the acceleration rate of the Universe in terms of the modified gravity theories continue to attract interest \cite{Joyce:2016vqv}. The simplest model which generalizes General Relativity is found by simply replacing the Ricci scalar ($R$) in the action by a function $f(R)$. This idea led to many modified gravity models studied in the literature  \cite{Nojiri:2017ncd} or in the cosmological set up of a higher-order modified teleparallel theory (see \cite{Lepe:2017yvs,Harko:2014aja,Otalora:2013dsa,Otalora:2013tba,Karpathopoulos:2017arc,Akarsu1,Akarsu2,Akarsu3,Akarsu4,Akarsu5} and references therein). Instead of doing modification of gravity, the nonlinear electrodynamics (NED) can be used to avoid singularities as well as resolve the horizon problem \cite{novello0}.

Recently, the idea of NED has been proposed as a solution to source the Universe acceleration \cite{kruglov1,kruglov2,kruglov3}.  In the early Universe the effect of the NED may have been very strong  and, in principle,  this may also explain the inflation.  In this scenario the NEF can be considered as a source of the gravitational field and, as a consequence, nonlinear magnetic fields may be a driven mechanism of the inflation of the Universe. In this line of research, very recently many NED models have been investigated using a stochastic magnetic background, i. e. the cosmic background with the wavelength smaller than the curvature, with a non-vanishing $< B^2 >$, where matter should be identified with a primordial plasma \cite{novello0,novello1,novello2,novello3,kruglov1, kruglov2,ao,Hendi:2012zz}. Thermal fluctuations in a dissipative plasma, i.e. plasma fluctuations, could source  stochastic magnetic fields on a scale larger than the thermal wavelength \cite{4}. Thus, there are the stochastic fluctuations of the electromagnetic field in a relativistic electron-positron plasma. Although homogeneous magnetic fields can affect  the isotropy of the Universe, i.e. the energy-momentum tensor can become anisotropic which could cause an anisotropic expansion law and modify the CMB spectrum, the effect on the Universe geometry (isotropy) of magnetic fields  tangled on scales much smaller than the Hubble radius are negligible \cite{4}. Thus, averaging the magnetic fields, which are sources in general relativity \cite{2}, give the isotropy of the Friedman-Robertson-Walker (FRW) spacetime. On the other hand, bulk viscosity term is neglected in the electric conductivity of the primordial plasma by taking $E^2=0$ \cite{tajima,gio,campos,dun,joy}.
The NED is useful to remove singularities of Big Bang and try to explain inflation naturally. Note that using the scalar fields for the inflation and early Universe have a problem with the fine exit. The problem is that after the inflation is started, it goes forever which is known as eternal inflation. However, we will propose a model where there is not any eternal inflation problem. For this purpose, we use the magnetic Universe with Nonlinear Electromagnetic Field (NEF) in the stochastic background with a nonzero value of $< B^2 >$ that supports the acceleration of the Universe. There are some cons and pros of NED from the Dirac-Born-Infeld (DBI) nonlinear electrodynamics such that for DBI theory, there is a duality  symmetry, on the other hand, for the NED, it is broken but it is also violated for the QED. Other difference is the birefringence phenomenon that occur in NED and also in QED with quantum corrections, but not in DBI theory \cite{novello2}. Moreover, the DBI model has a problem of causality.  

It is believed that magnetic fields have important role on the evolution of our universe. However, we have known little information about the existence of magnetic fields at the early universe. Observations have manifested the existence of magnetic field in the universe, ranging from the stellar scale $10^{-5}$ pc to the cosmological scale $10^4 \mathrm{Mpc}$ \cite{01,02}. In particular, the magnetic field on large scales ($\leq 1$ Mpc) is deemed to be formed in the early universe \cite{03}, namely, the primordial magnetic field. By the recent CMB observations, the strength of the magnetic field is less than a few nano-gauss at the 1 Mpc scale \cite{04}. Additionally, the $\gamma$-ray detections of the distant blazars imply that the magnetic field might be larger than $10^{-16}$ G on the scales $1 - 10^4$ Mpc \cite{05}. 

On the other hand, we have no direct observational evidence of primordial magnetic fields. The amplitude of primordial magnetic fields is also debatable. However, we believe that they existed because may have been needed to seed the large magnetic fields observed today. Nowadays, many theories are proposed to obtain the origin of cosmic magnetic fields for instance primordial vorticity plasma (vortical motion during the radiation era of the early Universe, vortical thermal background by macroscopic parity-violating currents), quantum-chromo-dynamics phase transition, first-order electroweak phase transition via a dynamo mechanism, etc (\cite{4} and references therein). In the last scenario, seed fields are provided by random magnetic field
fluctuations which are always present on a scale of the order of a thermal wavelength.
Primordial Nucleosynthesis limit the intensity of the magnetic seed fields to a current upper limit of $10^{-9}$ G \cite{3} and the lower limit $B > 10^{-19}$ G from the $\gamma$-ray observations \cite{44}. Today, magnetic fields have been observed in different types of galaxies and also cluster of galaxies at wide range of redshifts. Furthermore, a lower bound, $B \geq 3 \times 10^{-16}G$, has been obtained for intergalactic magnetic fields \cite{1}. Constraints from the Planck satellite in 2015 show that the upper limit to be of the order of  $B <10^{-9} G$ \cite{2}.

In late time epochs, the reason to use NED may be different than the early universe: it can be implemented as a phenomenological approach, in which the cosmic substratum is modeled as a material media with electric permeability and magnetic susceptibility that depend in nonlinear way on the fields \cite{10}. Another argument is based on the view that General Relativity is a low energy quantum effective field theory of gravity, provided that the Einstein-Hilbert classical action is augmented by the additional terms required by the trace anomaly characteristic of NED \cite{Montiel:2014dia}.

In this paper we use a new model of Lagrangian of NED, dubbed ``NED with an exponential correction'', which has a Maxwell limit at low energies and which is different than DBI nonlinear electrodynamics \cite{kruglov1}. We assume that radiation of NED is dominated in the early Universe, to solve the initial singularity problem. We show that the NED with the gravitation field can create the negative pressure and cause the inflation and late cosmic acceleration of the Universe. In this regard, our manuscript is an extended version of the papers 
\cite{ao,kruglov1,w1,Campanelli:2007cg,novello,novello2}, at which are used Nonlinear magnetic fields, as a source of inflation. This model of NED is valid for the early and current regime of the Universe, on the other hand for the late Universe, the magnetic field is very weak. However, the question whether one can explain the late regime of the Universe in terms of only NED remains, at least theoretically, as an open possibility. Hence, we additionally investigate whether the exponential NED can provide the late-time accelerated expansion of the Universe.

We perform a phase-space analysis of Einstein-NED cosmology without a matter source and including it. The advantage of the using phase-space analysis \cite{Leon:2012mt,Leon:2013qh,Kofinas:2014aka,Leon:2012vt,Fadragas:2013ina,Leon:2015via,Pulgar:2014cba,Leon:2014yua,Giacomini:2017yuk,Haba:2016swv,Stachowski:2016dfi,Hrycyna:2014cka,Szydlowski:2012zz,Szydlowski:2008in,Dabrowski:2003jm,Szydlowski:2008by,Szydlowski:2006ay,Godlowski:2005tw,Szydlowski:2005ph,gge} is that one can do more stability analysis with using visual plots using the trajectories in geometrical way so that it becomes easy to observe the property with the help of the attractors which are the most easily seen experimentally \cite{Marek1}. On the other hand, conceptually using NED has the advantage that no need to use some exotic fields such as scalar fields, branes or extra dimensions, it is just photon fields, and it is well known also in nonlinear optics  which studying behavior of light in nonlinear media and also nonlinear collision of particles in quantum electrodynamics \cite{DeLorenci:2000yh,Novello:1999pg,Novello:1979ik,Mourou:2006zz,DeLorenci:2001ch,VillalbaChavez:2012ea,Karbstein:2015xra}. 

The paper is organized as follows. In section \ref{Sect. 2} we introduce the Lagrangian of NED. In section \ref{Sec. 3} we examine the acceleration and evolution of the Universe in terms of NED fields. In section \ref{Sec. 4} we  perform a detailed phase-space analysis of our Einstein-NED cosmology model. In section \ref{Sec. 5} we consider a more realistic scenario, namely we include a matter source in our setup. In section \ref{Sec. 6} we put constraints on the NED parameters 
using observational Hubble data from cosmic chronometers and using the latest SNIa data. Finally we summarize and discuss our results in section \ref{Sec. 7}.

\section{General Relativity coupled with Non-Linear Electrodynamics with an exponential correction}
\label{Sect. 2}
In highly nonlinear energy density situations such as in the early Universe, NED is expected to play a crucial role in the evolution of the Universe \cite{w,magneticU1}. First it should be understand
the contributions of nonlinear fields to inflation. For this purpose,
we propose the following action GR coupled with the NED field as follows:
\begin{equation}
S=\int d^{4}x\sqrt{-g}\left[\frac{R}{2}+\mathcal{L}_{NED}\right],\label{Ac}
\end{equation}
where $R$ is the Ricci scalar
and $\mathcal{L}_{NED}$ is the Lagrangian of the NED fields and we have used units where $8 \pi G=1, c=1$. The new
NED Lagrangian density is chosen as follows:
\begin{equation}
{\cal L}_{NED}=-\frac{{\cal F}e^{-\alpha \cal F}}{(\alpha{\cal F}+\beta)},\label{1}
\end{equation}
where $\alpha$ is a constant with $[B_{0}^{-2}]$ units, and $B_0$ is the current value of the electromagnetic magnetic field,  and $\beta$ is a dimensionless parameter, ${\cal F}=(1/4)F_{\mu\nu}F^{\mu\nu}=(B^{2}-E^{2})/2$, where
$F_{\mu\nu}=\partial_{\mu}A_{\nu}-\partial_{\nu}A_{\mu}$ is the field
strength tensor. When $\alpha\rightarrow0$ and $\beta\rightarrow1$,
the Lagrangian reduces to that of the classical Maxwell's electrodynamics.

Notice that in geometrized units, where $8\pi G=1, c=1$, all the quantities have dimension of a power of length $[L]$. In this system of units, a quantity which has $L^{n}T^{m}M^{p}$ in ordinary units converse to $L^{n+m+p}$. To recover nongeometrized units, we have to use the conversion factor $c^{m}(8 \pi G/c^2)^{p}$. Thus, the dimension of $B_{0}$ and $H_{0}$ is [$L^{-1}$] in geometrized units and the conversion factors are $1 \text{Gauss}=1.44\times 10^{-24}\mathrm{cm}^{-1}$ and $H_{0}=h\,1.08\times 10^{-30}\mathrm{cm}^{-1}$, where $h =(73.24\pm1.74)\times 10^{-2}$ as measured by \cite{Riess:2016jrr}. Then we are dealing with magnetic fields of the order $10^{-40}\mathrm{cm}^{-1} \lesssim B_0 \lesssim 10^{-33}\mathrm{cm}^{-1}$ in the present epoch.

After varying the action given in Eq. (\ref{Ac}), one can find the Einstein field equations  and the NED fields equation as follows:
\begin{equation}
R_{\mu\nu}-\frac{1}{2}g_{\mu\nu}R=-T_{\mu\nu},\label{EQ}
\end{equation}
where
\begin{equation}
\partial_{\mu}\left(\sqrt{-g}\frac{\partial\mathcal{L}_{NED}}{\partial\mathcal{F}}F^{\mu\nu}\right)=0.
\end{equation}

The energy momentum tensor \cite{kruglov3}
\begin{equation}
T^{\mu\nu}=K_{\mu\lambda}F_{\nu}^{\lambda}-g^{\mu\nu}\mathcal{L}_{NED}
, \quad K_{\mu\lambda}=\frac{\partial\mathcal{L}_{NED}}{\partial\mathcal{F}}F_{\mu\lambda},\end{equation}
can be used to obtain the general form of the energy density $\rho_{NED}$
and the pressure $\ p_{NED}$ of NED fields as
\begin{align}
&\rho_{NED}=-{\cal \mathcal{L}}_{NED}-E^{2}\frac{\partial{\cal \mathcal{L}}_{NED}}{\partial{\cal F}}\nonumber\\
& =-{\frac{{{\rm e}^{-\alpha\,\cal F}}\left(\left({E}^{2}{\alpha}^{2}-\alpha\right){\cal F}^{2}+\beta\,\left({E}^{2}\alpha-1\right){\cal F}-{E}^{2}\beta\right)}{\left(\alpha\,{\cal F}+\beta\right)^{2}},}\label{rho}
\end{align}
and 
\begin{widetext}
\begin{align}
&p_{NED}={\cal \mathcal{L}}_{NED}+\frac{\left(E^{2}-2B^{2}\right)}{3}\frac{\partial\mathcal{L}_{NED}}{\partial\mathcal{F}}
\nonumber \\
&=-\frac{2}{3}\,{\frac{\left[\left(\left({B}^{2}-1/2\,{E}^{2}\right){\alpha}^{2}+3/2\,\alpha\right){\cal F}^{2}+\beta\,\left(3/2+\left({B}^{2}-1/2\,{E}^{2}\right)\alpha\right){\cal F}-\beta\,\left({B}^{2}-1/2\,{E}^{2}\right)\right]{{\rm e}^{-\alpha\,\cal F}}}{\left(\alpha\,{\cal F}+\beta\right)^{2}}}.\label{p}
\end{align}
\end{widetext}

To find the solution of the Einstein field equations, we consider
the homogeneous and isotropic cosmological metric of FRW with following
line element:
\begin{equation}
ds^{2}=-dt^{2}+a(t)^{2}\left[dr^{2}+r^{2}\left(d\theta^{2}+\sin^2\theta \,d\phi^{2}\right)\right]\label{eq:frwmetric},
\end{equation}
with scale factor $a(t)$.

The key point of this study is that it can be supposed that the stochastic
magnetic fields are the cosmic background with the wavelength smaller
than the curvature so we can use the averaging of EM fields which
are sources in GR and then we can obtain the isotropic FRW spacetime
\cite{tolman}. In general, the averaged EM fields have the properties:
\begin{equation}
\langle E\rangle=\langle B\rangle=0,\text{ }\langle E_{i}B_{j}\rangle=0,
\end{equation}
\[
\langle E_{i}E_{j}\rangle=\frac{1}{3}E^{2}g_{ij},\text{ }\langle B_{i}B_{j}\rangle=\frac{1}{3}B^{2}g_{ij},
\]
where the averaging brackets $\langle$ $\rangle$ is used for a simplicity.
The non-zero averaged magnetic field case is the most unexpected case
\cite{tolman}, where the magnetic field of the Universe is frozen
to occur the magnetic properties, it is necessary to screen the electric
field of the charged primordial plasma. We use the Eqs. (\ref{rho})
and (\ref{p}) (for $E^{2}=0)$ and obtain: 
\begin{equation}
\rho_{NED}={\frac{B^{2}{{\rm e}^{-1/2\,\alpha\,{B}^{2}}}}{\alpha\,{B}^{2}+2\,\beta}},\label{rhoo}
\end{equation}
and
\begin{equation}
p_{NED}=-{\frac{2{B}^{2}\left({B}^{4}{\alpha}^{2}+2\,\left(\beta+3/4\right)\alpha\,{B}^{2}-\beta\right){{\rm e}^{-\frac{\alpha{B}^{2}}{2}}}}{3\left(\alpha\,{B}^{2}+2\,\beta\right)^{2}}}.\label{p0}
\end{equation}
Through the analysis, it is imposed the energy condition $\rho_{NED}\geq 0$, which implies $2 \beta + \alpha B^2\geq 0$.

Afterwards, we use the\ FRW metric (\ref{eq:frwmetric}) and find
the equation of the Friedmann: 
\begin{equation}
3\frac{\ddot{a}}{a}=-\frac{1}{2}\left(\rho_{NED}+3p_{NED}\right),\label{Feqn}
\end{equation}
where we use the dot \char`\"{}.\char`\"{} to denote the time
derivative. Furthermore, we check the condition of the accelerated
Universe $\rho_{NED}+3p_{NED}<0$ with the sources of NED fields using
into Eqs. (\ref{rho}) and (\ref{p}):
\begin{align}
& \rho_{NED}+3p_{NED}= \nonumber\\
& -{\frac{2{B}^{2}{{\rm e}^{-\frac{\alpha{B}^{2}}{2}}}\left({B}^{4}{\alpha}^{2}+2\,\beta\,\alpha\,{B}^{2}+\alpha\,{B}^{2}-2\,\beta\right)}{\left(\alpha\,{B}^{2}+2\,\beta\right)^{2}}}.\label{100}
\end{align}

\begin{figure}[!ht]
\centering
\includegraphics[width=0.4\textwidth]{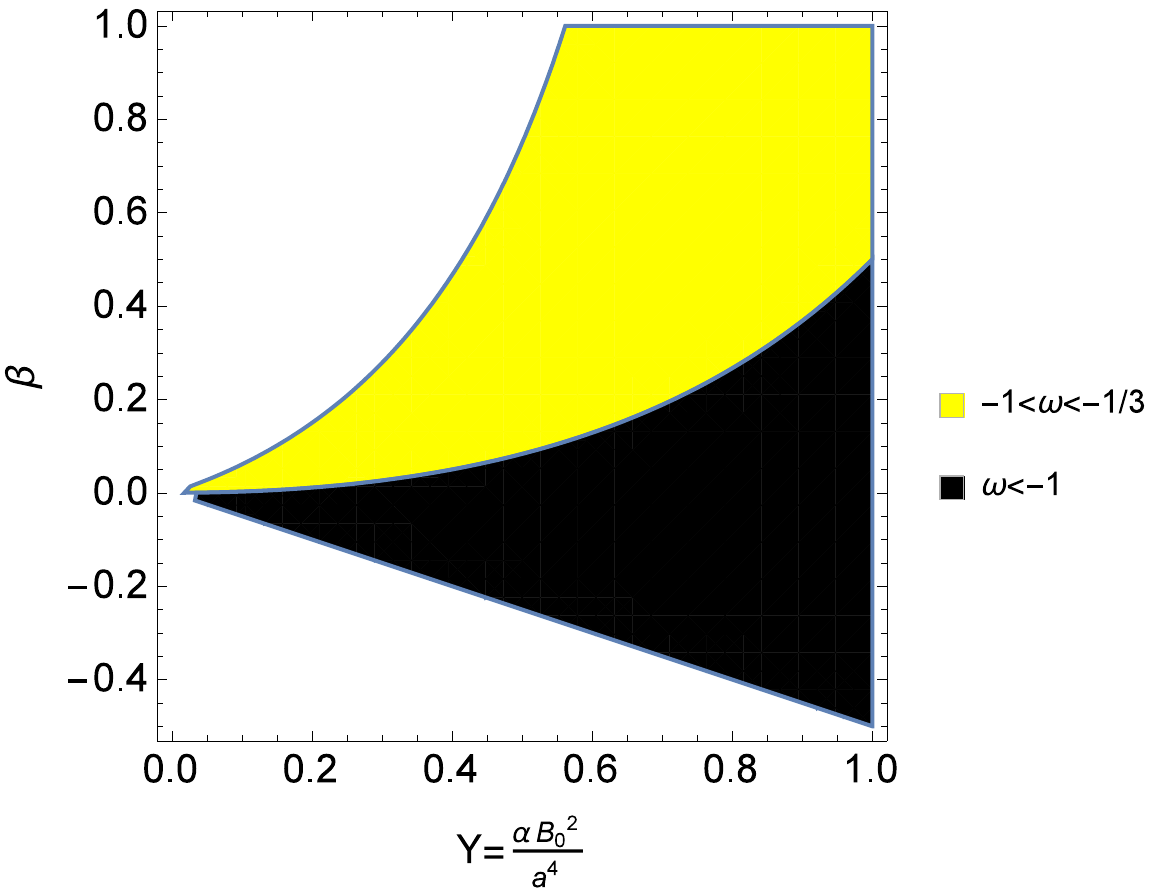}
\caption{\label{a10} Are presented the regions which gives acceleration in the plane $Y=\alpha B$ vs. $\beta$.}
\label{fig:yvsbeta}
\end{figure}

The condition for the NED energy provide acceleration is 
$(\rho_{NED}+3p_{NED}<0)$, which combined with the physical condition $\rho_{NED}\geq 0$ gives
$\beta \leq 0,\alpha\,B^{2}>-2 \beta$ or $\beta >0, \alpha\,B^{2}>-\frac{1}{2} (1+2 \beta)+\frac{1}{2} \sqrt{4 \beta ^2+12 \beta +1}$. As shown in Fig. (\ref{a10}) and the lower bound of $\alpha B$ that gives acceleration when $\beta=1$ is at $\alpha\,B^{2}=-\frac{3}{2}+\frac{1}{2}\sqrt{17}$. 
Note that the source of the strong nonlinear electrodynamics
field accelerates the Universe in the early stages, that is, for large $\alpha B^2$ we enter the shadowed region in Fig. \ref{a10}. Then
we use the conservation of the energy-momentum tensor $(\nabla^{\mu}T_{\mu\nu}=0$)
for the FRW metric (\ref{eq:frwmetric}) and find the continuity equation:
\begin{equation}
\dot\rho_{NED}+3\frac{\dot{a}}{a}\left(\rho_{NED}+p_{NED}\right)=0.\label{11}
\end{equation}
It is noted that the Hubble parameter is defined
as $H=\frac{\dot{a}}{a}$ which is the expansion rate of our Universe.
This equation gives
\begin{equation}
\frac{B e^{-\frac{1}{2} \alpha  B^2} \left(2 B H+\dot{B}\right)
   \left(-4 \beta +\alpha ^2 B^4+2 \alpha  \beta 
   B^2\right)}{\left(2 \beta +\alpha  B^2\right)^2}=0.
\end{equation}
One can integrate (one of the branches of) the above equation \footnote{We have also the trivial solution $B=0$, the regime $\alpha B^2 \rightarrow \infty$ and the constant solutions $B$, such that $ \left(-4 \beta +\alpha ^2 B^4+2 \alpha  \beta 
   B^2\right)=0$. We submit the reader to section \ref{Sect:obs_1} for a more complete discussion of this special cases.} using the $\rho$ and $p$ to
obtain the evolution of the magnetic field respect to the scale factor
$B(t)=\frac{B_{0}}{a(t)^{2}}$ where $B_{0}$ is for $a(t)=1$. Afterward,
we rewrite the energy density $\rho$ and the pressure $\ p$ using
the evolution of the magnetic field:
\begin{align}
&\rho_{NED}={\frac{{ B_{0}}^{2}}{{a}^{4}}{{\rm e}^{-{\frac{\alpha\,{ B_{0}}^{2}}{2{a}^{4}}}}}\left({\frac{\alpha\,{ B_{0}}^{2}}{{a}^{4}}}+2\,\beta\right)^{-1}},~~\label{rr}
\\
&p_{NED}=-\frac{2{ B_{0}}^{2}}{3{a}^{4}}{{\rm e}^{-{\frac{\alpha\,{ B_{0}}^{2}}{2{a}^{4}}}}} \left({\frac{\alpha\,{B_{0}}^{2}}{{a}^{4}}}+2\,\beta\right)^{-2} \times \nonumber \\
& \left({\frac{{ B_{0}}^{4}{\alpha}^{2}}{{a}^{8}}}+2\,{\frac{\alpha\,\left(\beta+3/4\right){ B_{0}}^{2}}{{a}^{4}}}-\beta\right). \label{pp}
\end{align}
Using the equation of state (EoS) 
\begin{align}
&\omega=\frac{p_{NED}}{\rho_{NED}}=-\frac{2}{3}\left({\frac{{\it B}^{4}_0{\alpha}^{2}}{{a}^{8}}}+2\,{\frac{\alpha\,\left(\beta+3/4\right){\it B}^{2}_0}{{a}^{4}}}-\beta\right)\times \nonumber\\
&\left({\frac{\alpha\,{\it B}^{2}_0}{{a}^{4}}}+2\,\beta\right)^{-1},
\end{align}
then we use Eqs. (\ref{rr}) and (\ref{pp}) and obtain the radiation
or other relativistic fluid case: 
\begin{equation}
\lim_{a\rightarrow\infty}\omega=\frac{1}{3}.\label{18-1}
\end{equation}

\begin{figure}[!ht]
\centering
\includegraphics[width=0.5\textwidth]{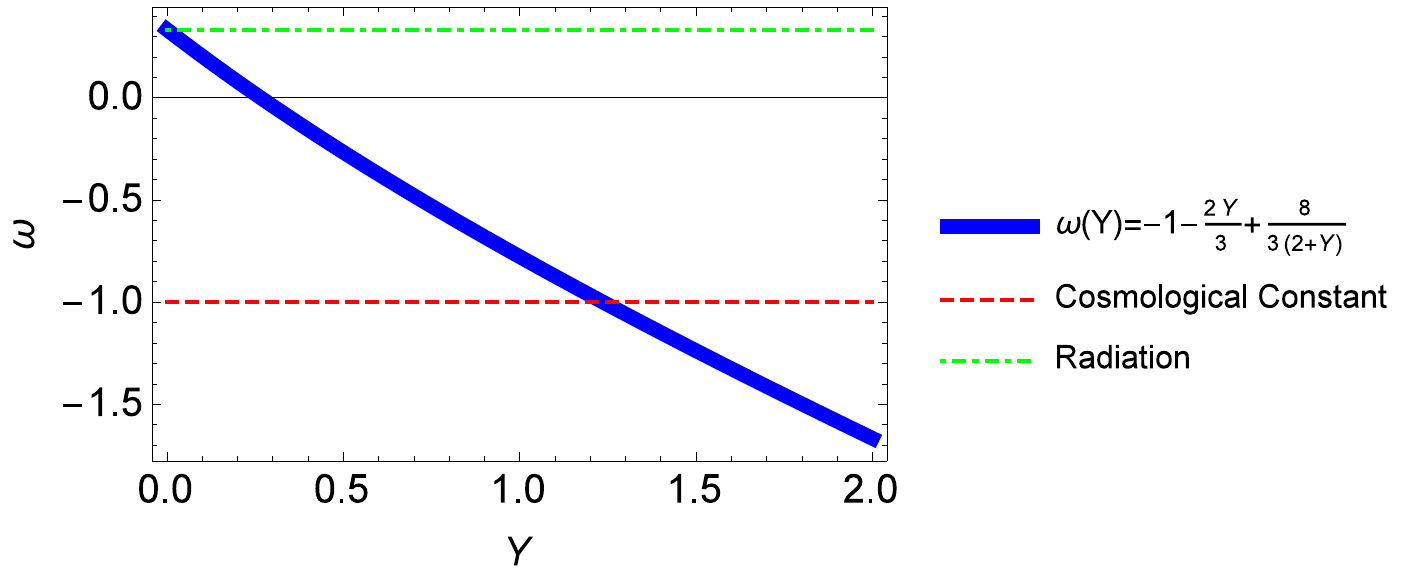}
\caption{\label{fig.11} $\omega$ versus $Y=\alpha B_{0}^{2}/a^{4}$ for $\beta=1$.}
\end{figure}

Defining $Y \equiv[\alpha B_{0}^{2}/a^{4}]$, we have $\omega=-\frac{2 \left(-\beta +Y^2+2 \left(\beta +\frac{3}{4}\right) Y\right)}{3 (2 \beta +Y)}$. 
Now, by choosing the value $\beta=1$, at $Y=\sqrt{5}-1$, is obtained an EoS corresponding to de Sitter spacetime i.e. $\omega=-1$ (the
usual cosmological constant), at $Y=\frac{65-\sqrt{7}}{4}$ , we have
$\omega=0$ for non-relativistic matter (baryons, CDM), also for some
values we have $-1<\omega<-1/3$ which is called ``quintessence'', a
dynamical dark energy resulting in accelerating expansion. 
The models with $\omega<-1$ has been termed ``phantom''
dark energy. We can also refer dark energy as vacuum energy, because
one possible source of dark energy is quantum-mechanical fluctuations
of the vacuum. In Fig. \ref{fig.11} is it plotted $\omega$ versus $Y=\alpha B_{0}^{2}/a^{4}$.
For small $a$ (large $Y$) the universe has large negative equation of state, it crosses the value $\omega=-1$ at $Y=\sqrt{5}-1$. At $Y= \frac{1}{2} \left(\sqrt{17}-3\right)$ the universe changes from acceleration to deceleration.  For large $a$ ($Y\rightarrow 0$) we see that $\omega$ approaches $1/3$ (radiation dominated universe).

Now we show that the spacetime will be flat at $t\rightarrow\infty$ ($a\rightarrow \infty$) and that singularities are removed at the early/late phase of the Universe ($a\rightarrow 0$). For this purpose we calculate the Ricci scalar (which represents the curvature of spacetime), the Ricci tensor squared, and the Kretschmann scalar. \\
The Ricci scalar is
calculated by using Einstein's field equation (\ref{EQ}) and the
energy-momentum tensor as follows:
\begin{equation}
R= (\rho_{NED}-3p_{NED}),
\label{16r}
\end{equation}
The Ricci tensor squared $R_{\mu\nu}R^{\mu\nu}$ and the Kretschmann
scalar $R_{\mu\nu\alpha\beta}R^{\mu\nu\alpha\beta}$ are also obtained
as 
\begin{equation}
R_{\mu\nu}R^{\mu\nu}=\left(\rho_{NED}^{2}+3p_{NED}^{2}\right),\label{rrrr}
\end{equation}
and 
\begin{equation}
R_{\mu\nu\alpha\beta}R^{\mu\nu\alpha\beta}=\left(\frac{5}{3}\rho_{NED}^{2}+2\rho_{NED} p_{NED}+3p_{NED}^{2}\right) \label{Kretschmann}.
\end{equation}
Furthermore, taking the limits $a\rightarrow0$ and at $a\rightarrow\infty$ in \eqref{rr} and in
\eqref{pp} we have \begin{align}
& \lim_{a\rightarrow0}\rho_{NED}(a)=\lim_{a\rightarrow0}p_{NED}(a)=0\\
& \lim_{a\rightarrow\infty}\rho_{NED}(a)=\lim_{a\rightarrow\infty}p_{NED}(a)=0.\label{15}
\end{align} 
Finally, when we check the limits of the energy density and pressure, and use the expressions \eqref{16r},   \eqref{rrrr} and \eqref{Kretschmann} we conclude
that the Ricci scalar, the Ricci tensor squared, and the Kretschmann scalar are non singular 
at $a(t)\rightarrow0$ and at $a(t)\rightarrow\infty$, and these show that the spacetime will be flat at $t\rightarrow\infty$ and singularities are removed at the early/late phase of the Universe.

\section{Acceleration and Evolution of the Universe}
\label{Sec. 3}
Now, we find the evolution of the Universe using the Einstein's equations
and the energy density given in Eqs. (\ref{rr}) and (\ref{pp}). Without
considering dust like matter, we study the evolution of the Universe. To calculate the scale factor as a function of time, first
we use the second Friedmann's equation in flat Universe:
\begin{equation}
\left(\frac{\dot{a}}{a}\right)^{2}=\frac{\rho_{NED}}{3},\label{fried}
\end{equation}
and one can obtain the equation which shows conservation of energy for
a particle moving in a effective potential $V(a)$:
\begin{equation}
\dot{a}^{2}+V_{\text{eff}}(a)=0,\label{eq:poten}
\end{equation}
with $V_{\text{eff}}(a)=-\frac{1}{6}\,{\frac{{B_{0}}^{2}}{{a}^{2}}{{\rm e}^{-\,{\frac{\alpha\,{B_{0}}^{2}}{2{a}^{4}}}}}\left(\,{\frac{\alpha\,{ B_{0}}^{2}}{{2a}^{4}}}+1\right)^{-1}}$.

\begin{figure}[!ht]
\includegraphics[width=0.3\textwidth]{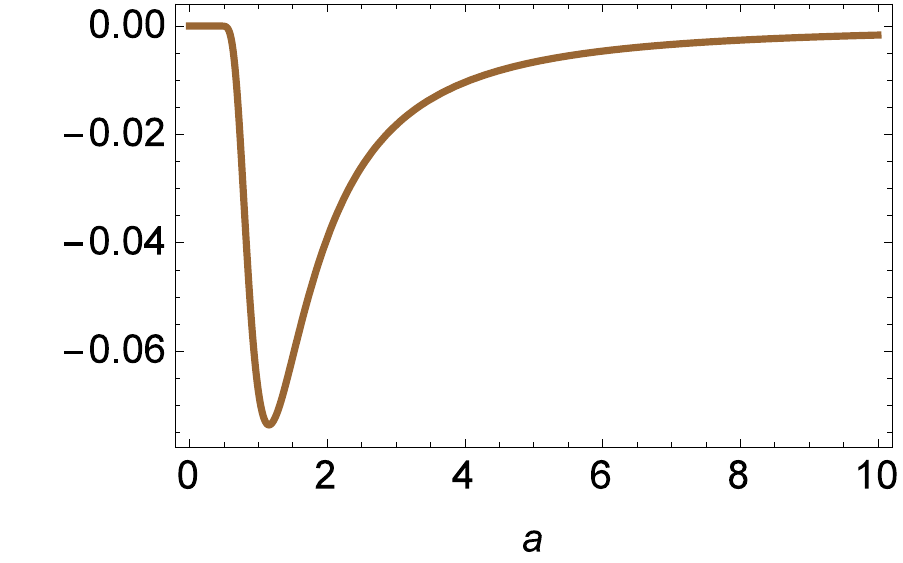}
\caption{\label{fig.1}The function of effective potential $V_{\text{eff}}(a)$ vs. scale factor $a$. }
\end{figure}

\begin{figure}[!ht]
\centering
\includegraphics[width=0.3\textwidth]{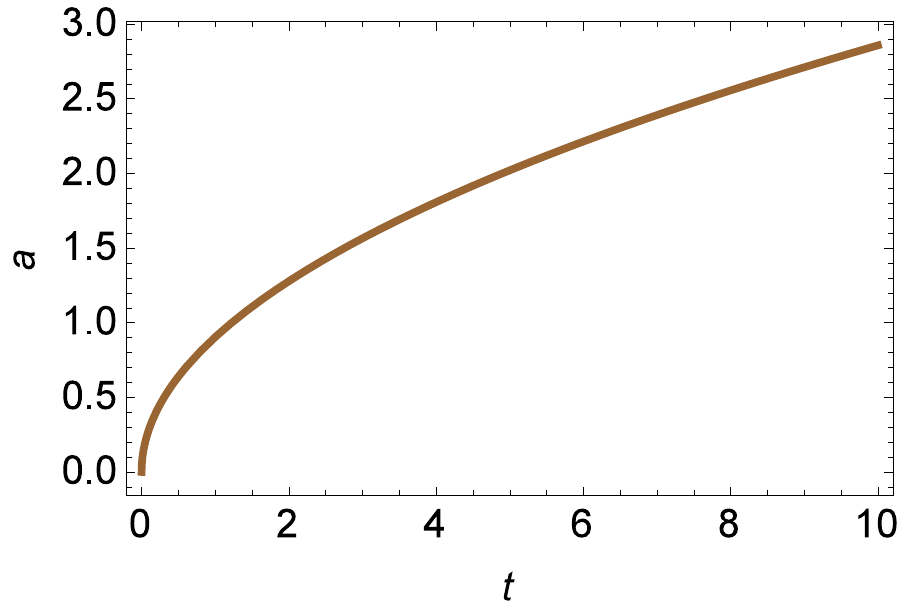}
\caption{\label{fig.33} The function of scale factor  $a(t)$ vs. cosmic time $t$. }
\end{figure}
We use equation (\ref{eq:poten}), to get a qualitative feel for the
evolution of the early Universe. Figure \ref{fig.1} shows the effective potential
$V_{\text{eff}}(a)$ as function of the scale factor.
The effective potential with a positive slope yields a force tending
to slow down positive motion along the horizontal axis, while the
portion of the effective potential with a negative slope yields a
force tending to speed up positive motion along the horizontal axis.
These two conditions occur to the right and the left, respectively,
of the minimum scale factor at $a_{c}=\frac{\sqrt{2}\sqrt[4]{\left(2\,\beta+1+\sqrt{4\,{\beta}^{2}+12\,\beta+1}\right){B_{0}}^{2}\alpha\,{\beta}^{3}}}{2\beta}\approx 1.15$. By analogy, then, $a(t)$
accelerates to the left of $a_{c}$ and decelerates to the
right of $a_{c}$. This acceleration is due to nonlinear
electrodynamics which behaves similarly to dark energy.

Using the Eq. (\ref{fried}) with the
energy density in Eq. (\ref{rr}), we obtain the equation as follows:
\begin{equation}
{ \dot{a}}^{2}=\frac{B_{0}^2}{6}  a^{-2} \left(\frac{\alpha  B_{0}^2}{2 a^4}+\beta \right)^{-1} e^{-\frac{\alpha  B_{0}^2}{2 a^4}}.\label{eq:a1}
\end{equation}
Expanding  (\ref{eq:a1})  around small $B=\frac{B_0}{a^2}$ (that is, large $a$), we obtain $\dot a^2=\frac{B_0^2}{6 a^2 \beta}+O\left(\frac{B_0}{a^2}\right)^3$. Neglecting error terms,  we calculate approximate cosmic time \cite{Balbi:2013dza}
as follows: $\frac{d a}{d t}=\frac{B_0}{\sqrt{6 \beta} a} \implies t-t_0=\frac{\sqrt{\frac{3}{2}} a^2 \sqrt{\beta }}{B_0}$ where $t_{0}$ is a constant of integration which gives only the shift
in time as shown in Fig. (\ref{fig.33}). Note that without losing generality, the integration constant has been shifted to $t_{0}=0$.  Then, by simplicity we have imposed the condition $\beta=1$. 
 One can also obtain the evaluation of the scale factor for ($\alpha=B_{0}=1)$ as
follows:
\begin{equation}
\label{eq(30)}
a=\sqrt[4]{2/3}\sqrt{\it t}.
\end{equation}

Note that evolution of the scale factor has similar feature with radiation dominated Universe  
and for $t=0$ it reduces to zero
\begin{equation}
a_{0}=a(t=0)=0.\label{phase}
\end{equation}

The function of $a_{0}$ is a radius of the Universe which shows that
Universe begins from the zero point. One can also show them in terms of redshift where $a=(1+z)^{-1}$. 

 Now, looking at large $a$ in figure \ref{fig.1} we see that $V_{\text{eff}}$ tends to zero as $a\to \infty$ (that, is, when $B=B_0 a^{-2}$ goes to zero); In figure \ref{fig.33}, $a$ goes to $\infty$. $\dot a \rightarrow \frac{B_0}{\sqrt{6} a \sqrt{\beta }}=\frac{\sqrt{B_0}}{2^{3/4} \sqrt[4]{3} \sqrt[4]{\beta } \sqrt{t-t_0}}$. All is consistent with $a\rightarrow \infty, \dot a\rightarrow 0, t\rightarrow \infty$. 

To complete the analysis we expand around large $B$, that is, small $a$, we have 
$\dot a^2= e^{-\frac{\alpha  B^2}{2}} \left(\frac{a^2}{3 \alpha }+O\left(\left(\frac{1}{B}\right)^2\right)\right)$. Substituting definition of $B=B_0 a^{-2}$, and neglecting error terms we obtain $\dot a^2 =\frac{a^2 e^{-\frac{\alpha  B_0^2}{2 a^4}}}{3 \alpha }$. Integrating we obtain $t-t_0=-\frac{1}{4} \sqrt{3} \sqrt{\alpha } \text{Ei}\left(\frac{B_0^2 \alpha }{4 a^4}\right)$. For $\alpha=B_0=1$ we have that $(\dot a, t-t_0)=\left(\frac{a e^{-\frac{1}{4 a^4}}}{\sqrt{3}},-\frac{1}{4} \sqrt{3} \text{Ei}\left(\frac{1}{4 a^4}\right)\right)$ that tends to $(0,-\infty)$ in the limit $a\rightarrow 0$. This solution corresponds to the shallow region of the potential showed in (\ref{fig.1}) near the origin  $a=0$, where $\dot a=-V_{\text{eff}}\rightarrow 0$, and it is achieved in the early universe. 

The next step is to check the causality of the Universe using the speed of the sound. The speed of the sound should be smaller than the local light speed ($c_{s}\leq1$) \cite{sound}. Moreover, we check the square sound
speed to avoid the Laplacian instability, we require the conditions that must be positive ($c_{s}^{2}>0$).Furthermore there exists a range in the parameter space where it is violated the condition for the absence of Laplace instability, "forbidden region".
Now our model is free of Laplacian instability.
Our all calculations are based on allowed region where there is no Laplacian instability.

\begin{figure}[!ht]
\vspace{-1cm}
\centering
\includegraphics[height=4in,width=4in,angle=2]{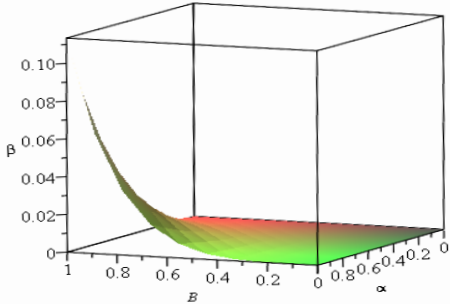}
\vspace{-4cm}
\caption{\label{fig.3}The plot shows the classical stability with variables $\alpha$, $\beta$ and $B$. }

\end{figure}

\begin{figure}[!ht]
\vspace{-1cm}
\centering
\includegraphics[height=4in,width=4in]{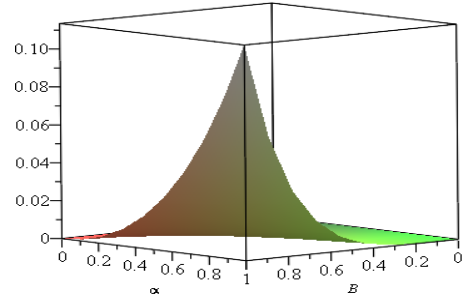}
\vspace{-4cm}
\caption{\label{fig.4}The plot shows the causality with variables $\alpha$, $\beta$ and $B$. }
\end{figure}

If the Universe satisfied these conditions, there is a classical stability. The square
of the sound speed is found from Eq.s (\ref{rho}) and (\ref{p});
\begin{widetext}
\begin{equation}
c_{s}^{2}=\frac{dp}{d\rho}=\frac{dp/d\mathcal{F}}{d\rho/d\mathcal{F}}
={\frac{-24\,{a}^{16}{\beta}^{2}-20\,\beta\,\left(\beta+{\frac{11}{5}}\right)\alpha\,{B_0}^{2}{a}^{12}+8\,\left(\beta-1/2\right)\beta\,{\alpha}^{2}{ B_0}^{4}{a}^{8}+8\,{\alpha}^{3}{ B_0}^{6}\left(\beta+3/8\right){a}^{4}+2\,{ B_0}^{8}{\alpha}^{4}}{24\,{a}^{16}{\beta}^{2}-12\,{ B_0}^{2}\beta\,\alpha\,\left(\beta-1\right){a}^{12}-12\,{ B_0}^{4}{a}^{8}{\alpha}^{2}\beta-3\,{ B_0}^{6}{a}^{4}{\alpha}^{3}}}, \label{cs1}
\end{equation}
\end{widetext}
and the classical stability $(c_{s}^{2}>0)$ for the absence of Laplacian instabilities and causality of the Universe using the speed of sound $c_{s}\leq1$, occurs as shown in Fig. \ref{fig.3} and Fig. \ref{fig.4}, respectively.

Finally, we deal with the configuration  ${\cal F} = - \beta/\alpha,$ which corresponds to $2 \beta + \alpha B^2= 0$. 
Using the parametrization $B=B_0 a^{-2}$, and the definition $Y=\alpha B_0^2 a^{-4}$, this configuration appears when $Y=-2 \beta$ (in Fig 2. $\beta$  has been chosen equal to $1$), and it is attained as $a\rightarrow a_s:= \frac{\sqrt[4]{-\alpha } \sqrt{B_0}}{\sqrt[4]{2}}$ ($\alpha$ negative). Expanding eq.  (\ref{eq:a1}) in a neighborhood of the value $a=a_s$ we have $\dot a = \frac{{B_0}^{3/4} \sqrt{e}}{2\ 2^{3/8} \sqrt{3} \sqrt[8]{-\alpha } \sqrt{a-a_s}}+O\left(\sqrt{a-a_s}\right)$. Hence, $a(t)\to a_s+\left(\frac{3}{2}\right)^{2/3} \sqrt[3]{C^2 t^2}$ where $C=\frac{{B_0}^{3/4} \sqrt{e}}{2\ 2^{3/8} \sqrt{3} \sqrt[8]{-\alpha }}$. The Hubble factor as $t\rightarrow 0$ goes as $H\rightarrow\frac{\sqrt[3]{\frac{2}{3}} \sqrt[3]{C^2}}{a_s} t^{-\frac{1}{3}}$. By integrating the approximated equation $\frac{\dot a}{a}\approx\frac{\sqrt[3]{\frac{2}{3}} \sqrt[3]{C^2}}{a_s} t^{-\frac{1}{3}}$ we have $a=a_s e^{\frac{\left(\frac{3}{2}\right)^{2/3} C^{2/3}
   t^{2/3}}{a_s}}=a_s+\left(\frac{3}{2}\right)^{2/3} C^{2/3} t^{2/3}+\frac{3
   \sqrt[3]{\frac{3}{2}} C^{4/3} t^{4/3}}{4a_s}+O\left(t^{5/3}\right)$, such that we obtain correction terms to the previous asymptotic formula as $t\rightarrow 0$.   On the other hand, for large $t$ we have $H\rightarrow \frac{2}{3} t^{-1}$, which corresponds to matter domination. The scale factor around $a_s$ corresponds to the inflationary solution $
a(t) = \exp[A t^f
]$ with $f=\frac{2}{3}$, that is followed with matter domination. A scale factor of the form $a(t) = \exp\big(At^f\big)$ where $A>0$ and $0<f<1$ was introduced in 
\cite{Barrow1990a, BarrowSaich1990, BarrowLiddle1993} in the context of inflation. Since the expansion of the universe with this scale factor is slower than the de Sitter inflation ($a(t) = \exp
(Ht)$ where $H$ is constant), but faster than the power-law
inflation ($a(t) = t^q$ where $q> 1$), it was called intermediate inflation. 
Intermediate inflationary models arise in the standard inflationary framework as exact cosmological solutions in the slow-roll approximation to
potentials that decay with inverse power-law of the inflaton field \cite{BarrowNunes2007}. These models have been studied in some warm inflationary scenarios  \cite{CampoHerrera2009, CampoHerrera2009a, HerreraVidela2010,  CidCampo2011, CidCampo2012, HerreraOlivaresVidela2013, HerreraOlivaresVidela2013a, Campo2014, HerreraOlivaresVidela2014, HerreraVidelaOlivares2015,Cid:2015pja}. From a quantum mechanical point of view, this mechanism might provide an explanation for the large scale magnetic fields observed today. In particular the inflation period amplifies the quantum perturbations of the electromagnetic field leading to the current  classical perturbations. 

\section{Phase space analysis}
\label{Sec. 4}
The equation \eqref{eq:poten} represents the motion of a particle of the unit mass in the potential
$U_{\text{eff}}(a) = -\frac{\rho_{NED}(a) a^2}{6}$. 
We have that the equation \eqref{eq:poten} is satisfied
on the zero energy level, where $\rho_{NED}$ plays the role of effective energy density parameterized
through the scale factor $a(t)$.  Therefore the standard cosmological
model can be simply represented in the terms of a dynamical system of a
Newtonian type: 
\begin{equation}
\ddot a=-\frac{\partial U_{\text{eff}}}{\partial a}, \quad U_{\text{eff}}=-\frac{B_0^2 a^2 e^{-\frac{\alpha  B_0^2}{2 a^4}}}{6
   \left(2 \beta  a^4+\alpha  B_0^2\right)},
\end{equation} 
where the scale factor $a$ plays the role of a positional
variable of a fictitious particle of the unit mass, which mimics the expansion of the
Universe.

For simplicity, we use introduce the new constant $b_0=\frac{1}{2} {B_0}^2 \alpha$, and introduce the time rescaling $\tau=\frac{t}{B_0}, B_0> 0$, and the variables 
\begin{equation}
x=a, \quad y=\frac{\dot a}{B_0},
\end{equation}

The system is then equivalent to 
\begin{align}
&\frac{d x}{d\tau}= y, \quad 
\frac{d y}{d \tau}= \frac{e^{-\frac{b_0}{x^4}} \left(2 b_0^2-\beta  x^8+b_0 x^4 (2\beta +1)\right)}{6 x^3 \left(b_0+\beta 
   x^4\right)^2}.
\end{align}

By definition $x\geq 0$ (since we consider that the scale factor is non-negative). For this reason, it is convenient to define the new variables 

\begin{equation}
x=e^u, y=v,  
\end{equation}
which takes values on the real line, and the time variable
\begin{equation}
\eta=\int a^{-1} d\tau\equiv \int e^{-u} d\tau.
\end{equation}

This system can be written in the form 
\begin{align}
&\frac{d u}{d\eta}= v, 
\quad \frac{d v}{d \eta}= -\frac{\partial W(u)}{\partial u}. 
\end{align}

Thus,  $\frac{v^2}{2}+W(u)=E$, is the constant of energy. 
From the above system we see that, generically, the fixed points 
are situated on the axis $u$ ($v=0$). From the characteristic equation it follows that just three types of fixed points are admitted: 

\begin{enumerate}
\item saddle if $u_0: \frac{\partial W}{\partial u}|_{u=u_0}=0$ and $\frac{\partial^2 W}{\partial u^2}|_{u=u_0}<0;$ 
\item focus if $u_0: \frac{\partial W}{\partial u}|_{u=u_0}=0$ and $\frac{\partial^2 W}{\partial u^2}|_{u=u_0}>0;$ 
\item degenerated critical point if  $u_0: \frac{\partial W}{\partial u}|_{u=u_0}=0$ and $\frac{\partial^2 W}{\partial u^2}|_{u=u_0}=0.$ 
\end{enumerate}

In the concrete example we have 
\begin{equation}
W(u)=-\frac{e^{-b_0 e^{-4 u}}}{12 \left(b_0 e^{-2 u}+\beta  e^{2 u}\right)}.
\end{equation}

Imposing the condition $b _{0}+\beta  e^{4 u}\geq 0$ we find that the fixed point at the finite region of the phase space is
$A:(u, v)=\left(\frac{1}{4} \ln \left(\frac{b_{0} \left(1+2 \beta +\sqrt{4 \beta  (\beta +3)+1}\right)}{2 \beta }\right), 0\right)$, that exists for $\beta >0, b_{0}>0$. Then,  $W''(u_0)$ is positive  (where $u_0$ is the coordinate of A).

So that, according to the previous classification it is a focus. 
 For the particular case $b_0=0.5, \beta=1$ we find that the fixed point $A$ has coordinates $u=0.144262$ for which $a=1.15519$ which corresponds to the minimum of $V(a)$ in Fig \ref{fig.1}. Additionally, in the limit $u\rightarrow + \infty$, we have $W'(u)\rightarrow 0, W''(u)\rightarrow 0$, hence, in this regime we can expect to have degenerate critical points. Furthermore, for ${b_0}{\beta}<0$ there are no fixed points  at the finite region of the phase-plane.

   Since the above system is in general unbounded, then we introduce the compactification 
 \begin{equation}
 U=\frac{u}{\sqrt{1+u^2+v^2}}, \quad V=\frac{v}{\sqrt{1+u^2+v^2}},
 \end{equation}
 we have the equivalent flow 
\begin{widetext}
 \begin{align}
 & \frac{d U}{d\eta}=\frac{U V \sqrt{1-U^2-V^2}  \left(-2 b_{0}^2-b_{0} (2 \beta +1) e^{\frac{4 U}{\sqrt{1-U^2-V^2}}}+\beta  e^{\frac{8 U}{\sqrt{1-U^2-V^2}}}\right) \exp \left(-b_{0} e^{-\frac{4 U}{\sqrt{1-U^2-V^2}}}-\frac{2 U}{\sqrt{1-U^2-V^2}}\right)}{6 \left(b_{0}+\beta  e^{\frac{4
   U}{\sqrt{1-U^2-V^2}}}\right)^2} \nonumber \\
   & \;\;\;\;\;\;\;\;\;\;\;\;\;\; +  V(1-U^2-V^2)+V^3, \label{Eq-4.15}\\
 &\frac{d V}{d\eta}=\frac{\left(V^2-1\right) \sqrt{1-U^2-V^2} \left(-2 b_{0}^2-b_{0} (2 \beta +1) e^{\frac{4 U}{\sqrt{1-U^2-V^2}}}+\beta  e^{\frac{8
   U}{\sqrt{1-U^2-V^2}}}\right)  \exp \left(-b_{0} e^{-\frac{4 U}{\sqrt{1-U^2-V^2}}}-\frac{2
   U}{\sqrt{1-U^2-V^2}}\right)}{6 \left(b_{0}+\beta  e^{\frac{4 U}{\sqrt{1-U^2-V^2}}}\right)^2}\nonumber \\
 & \;\;\;\;\;\;\;\;\;\;\;\;\;\; -U V^2. \label{Eq-4.16}
 \end{align}
\end{widetext}

\begin{figure}[!ht]
\centering
\includegraphics[width=0.3\textwidth]{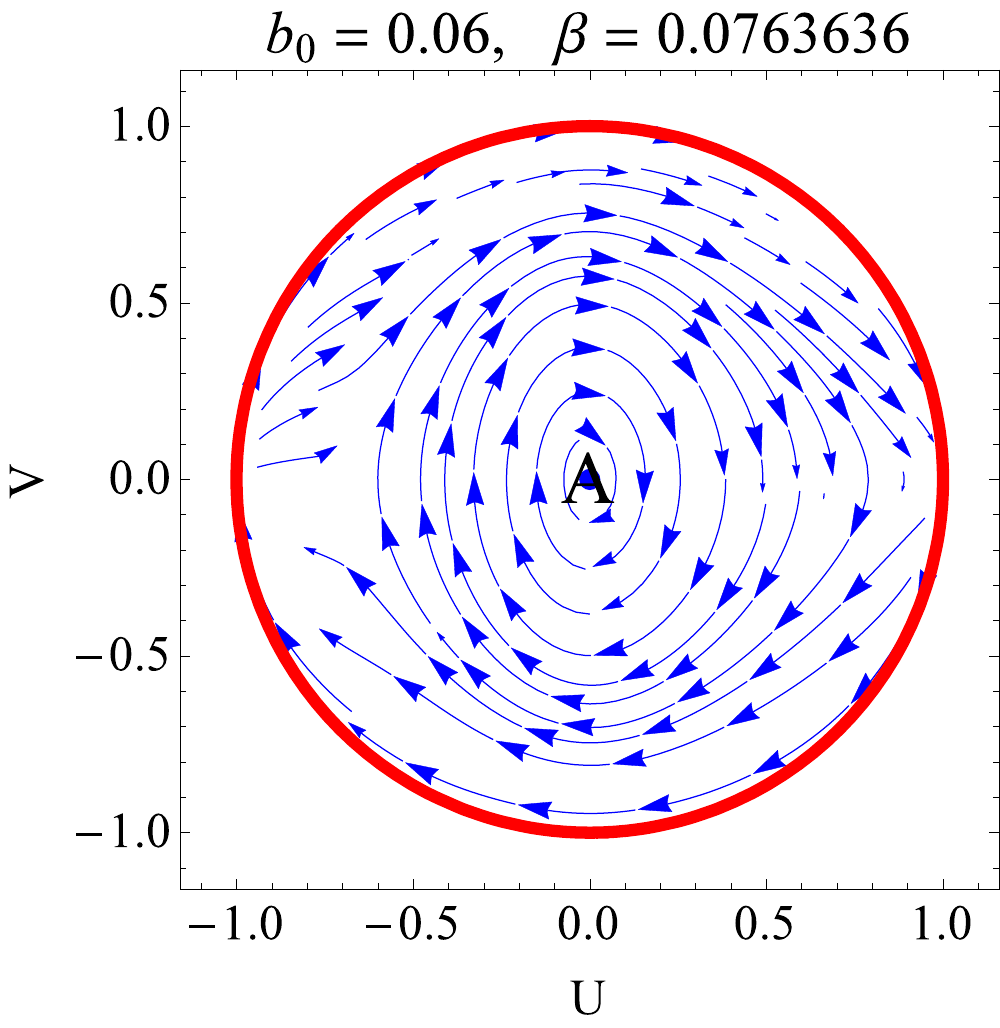}
\caption{\label{fig5} Dynamics of system 
\eqref{Eq-4.15}-\eqref{Eq-4.16} for the choice $\beta= \frac{2 b_{0}^2+b_{0}}{1- 2 b_{0}}, b_{0}=0.06$.}
\end{figure}

For the choice $\beta= \frac{2 b_{0}^2+b_{0}}{1- 2 b_{0}}, 0<b_{0}<\frac{1}{2}$ which implies $\beta >0, b_0>0$,  the fixed point $A$ exists.
In Fig. \ref{fig5} it is presented the dynamics of system 
\eqref{Eq-4.15}-\eqref{Eq-4.16} for the choice $\beta= \frac{2 b_{0}^2+b_{0}}{1- 2 b_{0}}, b_{0}=0.06$.
In this case $A$ is a focus according to the previous classification. 
Now, for $\beta= \frac{2 b_{0}^2+b_{0}}{1- 2 b_{0}}, 0<b_{0}<\frac{1}{2}$, $\frac{e^{-b_{0}} \left(2 b_{0}\left(4 b_{0}^2-6 b_{0}+1\right)+1\right)}{6 b_{0}}=: k^2>0 $. Then, the equations near the origin of coordinates (fixed point $A$), can be written as
 \begin{equation}
 \frac{d U}{d\eta}=V, \quad
 \frac{d V}{d\eta}=-k^2 U
 \end{equation}
where we have neglected the higher order terms,
such that, close to the origin, the solutions can be approximated by 
 \begin{align}
& U(\eta )=U_0 \cos (\eta  k)+\frac{V_0 \sin (\eta  k)}{k}, \label{EQQ_41}\\
& V(\eta )= -k U_0 \sin (\eta  k)+V_0 \cos (\eta  k).  \label{EQQ_42}
 \end{align}
Thus, the orbits of the original system near the origin can be approximated by the ellipses given by 
 \begin{equation}
 U(\eta )^2+k^{-2}V(\eta )^2=\text{constant}=U_0^2+k^{-2}V_0^2,
 \end{equation}
which leads to periodic solutions:
\begin{equation}
a=e^{\frac{U}{\sqrt{1-U^2-V^2}}}, \quad \dot a=\frac{B_0 V}{\sqrt{1-U^2-V^2}},
\end{equation}
where $U$ and $V$ are given by \eqref{EQQ_41}, \eqref{EQQ_42}

\begin{figure}[!ht]
\centering
\includegraphics[scale=0.6]{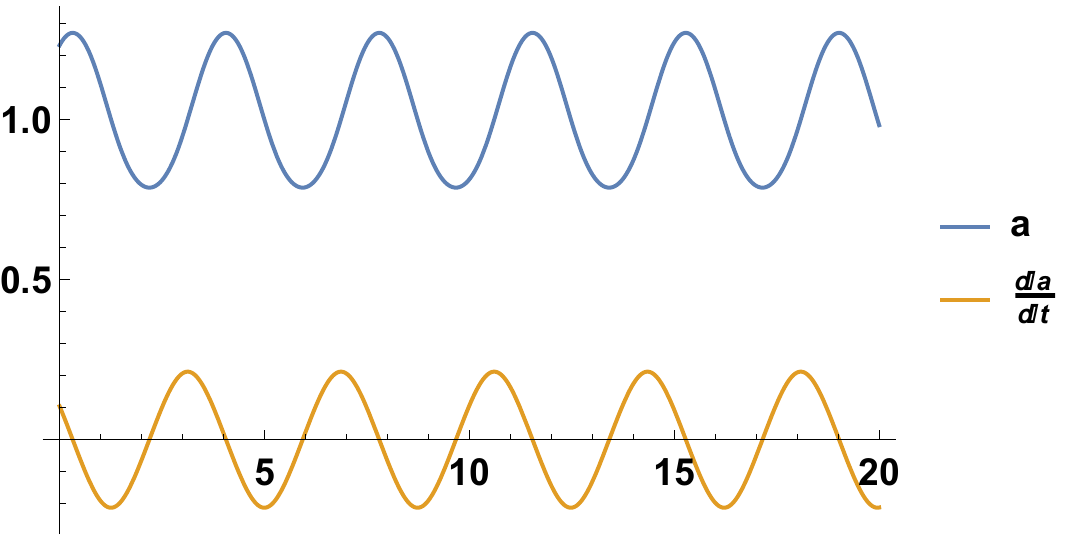}
\caption{\label{cyclic} Behavior of the scale factor and its first derivative in terms of the parameter $\eta$. In the figure 
it is shown a typical behavior  for the choice $\beta= \frac{2 b_{0}^2+b_{0}}{1- 2 b_{0}}, b_{0}=0.06$, $B_0=0.5, U_0=V_0=0.2$.}
\end{figure}

The relation between $\eta$ and the cosmic time $t$ can be found by using 
\begin{equation}
t-t_0=B_0 \int_0^{\eta} a(\vartheta) d \vartheta,
\end{equation}
which in general must be integrated numerically.

In the figure \ref{cyclic} it is shown the behavior of the scale factor and its first derivative in terms of the parameter $\eta$. This is the realization of a cyclic Universe in our model supported by NED. The scale factor goes below and above the value $a=1$, reaching a maximum and a minimum value of $a$, and $a$ is bounded away zero (there is no initial singularity, as expected from our NED proposal). 

Now let us assume arbitrary $b_0>0, \beta>0$ and obtain a second order expansion around $A$. For the sake of simplicity of notation we define $\mu= \frac{b_{0} \left(1+2 \beta\sqrt{4 \beta ^2+12 \beta +1}\right)}{2 \beta }$ and $k^2:=\frac{e^{-\frac{b_{0}}{\mu }} (\mu -2 b_{0}) \left(-4 b_{0}^2+4 b_{0} \mu +\mu ^2\right) \left(\ln ^2(\mu )+16\right)}{96
  b_{0} \mu ^{5/2}}>0$. The existence conditions $b_0>0, \beta>0$ implies $b_{0}>0, \mu >2 b_{0}$.
 
 By neglecting higher orders terms we obtain 
    \begin{align}
 & \frac{d U}{d\eta}=V \left(1-\frac{\ln ^2(\mu )}{\ln ^2(\mu )+16}\right)\\
 &\frac{d V}{d\eta}=-k^2 \left(U-\frac{\ln (\mu )}{\sqrt{\ln ^2(\mu )+16}}\right).
 \end{align}

Taking as initial condition $U(0)=\delta_{U}+\frac{\ln (\mu )}{\sqrt{\ln ^2(\mu )+16}},V(0)=\delta_{V}$, and under the conditions $b_0>0, \beta>0$,
the solution is given by 
\begin{align}
 & U(\eta )=\frac{\ln (\mu )}{\sqrt{\ln ^2(\mu )+16}}+ \delta_{U} \cos \left(\frac{4 \eta  k}{\sqrt{\ln ^2(\mu )+16}}\right) \nonumber\\
 & +\frac{4 \delta _{V} \sin \left(\frac{4 \eta  k}{\sqrt{\ln ^2(\mu
   )+16}}\right)}{k \sqrt{\ln ^2(\mu )+16}}, \label{Eq:4.23}\\
& V(\eta )=-\frac{1}{4} \delta_{U} k \sqrt{\ln ^2(\mu )+16} \sin \left(\frac{4 \eta  k}{\sqrt{\ln ^2(\mu )+16}}\right) \nonumber \\
&+\delta_{V} \cos \left(\frac{4 \eta  k}{\sqrt{\ln ^2(\mu
   )+16}}\right), \label{Eq:4.24}
\end{align}

This solution approximates the exact solutions of the full system surrounding $A$. The orbits near $A$ can be approximated by the ellipses 

\begin{align}
&\left(U-\frac{\ln (\mu )}{\sqrt{\ln ^2(\mu )+16}}\right)^2+\frac{16 V^2}{k^2 \left(\ln ^2(\mu )+16\right)}\nonumber \\& =\delta _{U}^2+\frac{16\delta_{V}^2}{k^2 \left(\log
   ^2(\mu )+16\right)}.
\end{align}

with center $A$. 
Once we know the expressions of $U(\eta), V(\eta)$, we can calculate the parametric expressions of $a$ and $\dot a$ as functions of $\eta$ through 
\begin{equation}
a=e^{\frac{U}{\sqrt{1-U^2-V^2}}}, \quad \dot a=\frac{B_0 V}{\sqrt{1-U^2-V^2}},
\end{equation}
where $U$ and $V$ are defined by \eqref{Eq:4.23} and \eqref{Eq:4.24}, respectively. 
Finally, the dynamics at the circle at infinity can be represented by the flow of 

 \begin{align}
 & \frac{d U}{d\eta}=V^3, \quad
\frac{d V}{d\eta}=-U  V^2.
 \end{align}

 The orbits lying on the circle at infinity can be parametrized as
 $$U(\eta)= \pm \frac{c_2+\eta }{\sqrt{\left(c_2+\eta\right){}^2+1}}, \quad V(\eta)= \sqrt{\frac{1}{\left(c_2+\eta \right){}^2+1}}$$
    or 
    $$U(\eta)=\pm \frac{\eta -c_2}{\sqrt{\left(\eta -c_2\right){}^2+1}}, \quad V(\eta )= -\sqrt{\frac{1}{\left(\eta -c_2\right){}^2+1}}.$$
  Thus, there are fixed points lying on the circumference at infinity  $(U,V)=(\pm 1,0)$. Thus, apart of the singular points at infinity described before, and the point at the finite region  $(U,V)=\left(\frac{\ln\left(\mu\right)}{\sqrt{\ln^2\left(\mu\right)+4}},0\right)$, $\mu= \frac{b_{0} \left(1+2 \beta\sqrt{4 \beta ^2+12 \beta +1}\right)}{2 \beta }$, we have that for $b_0 \beta <0$, there exists the singular line: 
  $\left\{(X,Y):b_{0}+\beta  e^{\frac{4 U}{\sqrt{1-U^2-V^2}}}=0\right\}.$

\subsection{Integrability and connection with the observables}
\label{Sect:obs_1}
 
In this section we comment on the integrability of the system at hand, and calculate some observables in terms of redshift. 

As we commented before, from the conservation of the energy-momentum tensor $(\nabla^{\mu}T_{\mu\nu}=0$)
for the FRW metric we have
\begin{equation}
\frac{B e^{-\frac{1}{2} \alpha  B^2} \left(2 B H+\dot{B}\right)
   \left(-4 \beta +\alpha ^2 B^4+2 \alpha  \beta 
   B^2\right)}{\left(2 \beta +\alpha  B^2\right)^2}=0,
\end{equation}
and from \eqref{rhoo} we have 
\begin{equation}
3H^2:=\rho_{NED}={\frac{B^{2}{{\rm e}^{-1/2\,\alpha\,{B}^{2}}}}{\alpha\,{B}^{2}+2\,\beta}} \label{eq:4.18DS}. 
\end{equation}
As we mentioned before, we have the trivial solution $B=0$, the regime $\alpha B^2\rightarrow \infty$ and the constant solutions $B$, such that $ \left(-4 \beta +\alpha ^2 B^4+2 \alpha  \beta 
   B^2\right)=0$. Concerning the former solutions we have the following results:
 \begin{enumerate}
   \item $\alpha B^2= -\beta -\sqrt{\beta^2 + 4\beta}$, is a real value for $\beta >0, \alpha <0 \;\text{or}\; \beta \leq -4, \alpha >0$. Equation \eqref{eq:4.18DS} means that $H$ is constant and equal to a cosmological constant $H=\sqrt{\Lambda}$, with $$\Lambda=\frac{e^{\frac{1}{2} \left(\beta +\sqrt{\beta  (\beta +4)}\right)} \left(\beta
   +\sqrt{\beta  (\beta +4)}+2\right)}{6 \alpha },$$ but under the above conditions results in $\Lambda<0$, which implies $a\propto e^{i \sqrt{|\Lambda|} t}$. So, we discard these solutions. Furthermore, we have a second class of solutions given by
   \item $\alpha B^2= -\beta +\sqrt{\beta^2 + 4\beta}$ is a real value for $\beta >0, \alpha >0 \;\text{or}\; \beta \leq -4, \alpha >0$. As before,  equation \eqref{eq:4.18DS} means that $H$ is constant and equal to a cosmological constant $H=\sqrt{\Lambda}$
   with 
   $$\Lambda=\frac{e^{\frac{1}{2} \left(\beta -\sqrt{\beta  (\beta +4)}\right)} \left(\beta
   -\sqrt{\beta  (\beta +4)}+2\right)}{6 \alpha }.$$
   For $\beta \leq -4, \alpha >0$ we have $\Lambda<0$, which implies $a\propto e^{i \sqrt{|\Lambda|} t}$. So, we discard these solutions.  However, for $\beta >0, \alpha >0$, we obtain $\Lambda>0$ as required in, and then we arrive a a regime similar to de-Sitter type Universe, with $a\propto e^{\sqrt{\Lambda} t}$.   
   \end{enumerate}
Summarizing, our model supports the inflation similar to de-Sitter Universe, with 
\begin{equation}
a\propto e^{\sqrt{\Lambda} t}, \quad \Lambda=\frac{e^{\frac{1}{2} \left(\beta -\sqrt{\beta  (\beta +4)}\right)} \left(\beta
   -\sqrt{\beta  (\beta +4)}+2\right)}{6 \alpha },
   \end{equation}
   provided  $\beta >0, \alpha >0$.  

Now, assuming that $B$ is not a constant or trivial, it is easy to recast the field equations as
\begin{equation}
\dot B= -2 B H, \quad \dot H= \frac{B^2 e^{-\frac{1}{2} \alpha  B^2} \left(-4 \beta +\alpha ^2 B^4+2 \alpha  \beta  B^2\right)}{3 \left(2 \beta +\alpha  B^2\right)^2}, \label{eee}
\end{equation}
The other restriction from Eq. (\ref{eq:4.18DS}) is
\begin{equation}
\label{rest_62}
\frac{B^2 e^{-\frac{1}{2} \alpha  B^2}}{2 \beta +\alpha 
   B^2}=3 H^2\geq 0.
\end{equation}

Using the previous restriction and introducing the logarithmic time variable $N=\ln a$,
we obtain the equations
\begin{equation}
\label{Eq_63}
\frac{d B}{d N}= -2 B, \quad \frac{d H}{d N}= \frac{H \left(-4 \beta +\alpha ^2 B^4+2 \alpha  \beta  B^2\right)}{2 \beta +\alpha  B^2},
\end{equation}
We use convention that $a=1$, is the present value of the scale factor, so that $N=0$ today. We define $H(0)=H_0$ and $B(0)=B_0$ the current values of the Hubble factor and the magnetic field respectively. It is convenient to define the dimensionless parameters
\begin{equation}
h_0=\alpha H_0^2, \quad b_0=\frac{\alpha B_0^2}{2}.
\end{equation}

Integrating the system \eqref{Eq_63}, evaluating $N=\ln a$ and $a=\frac{1}{1+z}$, we obtain
\begin{align}
& B(z)=B_0 (z+1)^2,\\
& H(z)=\frac{H_0 (z+1)^2 \sqrt{ \beta +b_0} e^{-\frac{1}{2} b_0\left((z+1)^4-1\right)}}{\sqrt{ \beta + b_0(z+1)^4}}, \label{eq:Hznomatter}\\
&q=-1- 2 b_0 (z+1)^4+\frac{2 \beta }{ \beta + b_0 (z+1)^4},
\label{eq:qznomatter}
\end{align}
where we have taken the reference values $z=0, a=1$ for today, such that, as $z\rightarrow -1$, the contribution of magnetic fields at cosmological scales is negligible \cite{kierdorf}.
The current value of the deceleration parameter is
\begin{align}
&q_0=1- 2\left(3 h_0 e^{b_0}+ b_0\right).
\end{align}
Thus, to accommodate the current accelerated phase it is required $h_0 >\frac{1}{6}e^{-b_0}\left(1-2 b_0\right)$.
Summarizing, the function $H(z)$ has the free parameters $H_0, h_0, b_0$ than can be contrasted against data and experiments. Then, we calculate the model parameters as:
\begin{equation}
\label{eq_74}
\alpha=\frac{h_0}{H_0^2},\; B_0=H_0 \sqrt{\frac{2 b_0}{h_0}},\; \beta=  b_0 \left(\frac{e^{-b_0}}{3h_0}-1\right).
\end{equation}

Notice that our analysis is performed in geometrized units where the dimension of $B_{0}$ and $H_{0}$ is [$L^{-1}$]. In geometrized units, $8\pi G=1$ and $c=1$, the conversion factors are $1 \text{Gauss}=1.44\times 10^{-24}\mathrm{cm}^{-1}$, and $H_{0}=h\,1.08\times 10^{-30}\mathrm{cm}^{-1}$. Then we are dealing with magnetic fields of the order $10^{-40}\mathrm{cm}^{-1} \lesssim B_0 \lesssim 10^{-33}\mathrm{cm}^{-1}$ in the present epoch \cite{1,2}. 

\section{General Relativity coupled with Non-Linear Electrodynamics with an exponential correction in the presence of matter}
\label{Sec. 5}
To introduce a more realistic model we include an additional matter source with constant equation of state parameter $w_m=p_m/\rho_m$ (with $0\leq w_m\leq 1$ for standard matter) and with continuity equation 
\begin{equation}
\dot \rho_m +3 H(1+w_m)\rho_m=0.
\end{equation}
Integrating out the last equation we obtain $\rho_m=\rho_{m,0} a^{-3(1+w_m)}$.
In this case, the second Friedmann's equation in flat Universe becomes 

\begin{equation}
\left(\frac{\dot{a}}{a}\right)^{2}=\frac{\rho_{NED}+\rho_{m,0} a^{-3(1+w_m)}}{3},\label{FRIED}
\end{equation}

Hence, the effective potential is now 

\begin{equation}
U_{\text{eff}}=-\frac{B_0^2 a^2 e^{-\frac{b_0}{a^4}}}{12
   \left( \beta  a^4+ b_0\right)}-\frac{\rho_{m,0}}{6} a^{-1-3 w_m}. \end{equation}
where, as in the previous case, we have introduced the new constants $b_0=\frac{1}{2} {B_0}^2 \alpha$ and $\rho_{m,0} =\rho_0 B_0^2$. Defining the variables 
\begin{equation}
a=e^u, \frac{\dot a}{B_0}=v,  
\end{equation}
which takes values on the real line, and the new time variable
\begin{equation}
\eta=\frac{1}{B_0}\int a^{-1} dt\equiv \frac{1}{B_0} \int e^{-u} dt,
\end{equation}
the system is then equivalent to 
\begin{align}
&\frac{d u}{d\eta}= v, \quad \frac{d v}{d \eta}= -\frac{\partial U_{\text{eff}}}{\partial u}.
\end{align}
where the effective potential is
\begin{equation}
U_{\text{eff}}(u)=-\frac{1}{6} \rho_{0} e^{-u (3 w_m+1)}-\frac{e^{-b_0 e^{-4 u}}}{12 \left(b_0 e^{-2 u}+\beta  e^{2 u}\right)}
\end{equation}

Now, the fixed points are found by solving numerically 
\begin{align}
\label{Px}
& \frac{e^{-b_{0} e^{-4 u}-2 u} \left(2 b_{0}^2+b_{0} (2 \beta +1) e^{4 u}-\beta  e^{8 u}\right)}{6 \left(b_{0}+\beta  e^{4
   u}\right)^2} \nonumber \\
	& \;\;\;\;\;\;\;\;\;\;-\frac{1}{6} \rho_{0} (3 w_m+1) e^{-u (3 w_m+1)}=0. 
\end{align}

As before the above system is in general unbounded, so that we introduce the compactification 

 \begin{equation}
 U=\frac{u}{\sqrt{1+u^2+v^2}}, \quad V=\frac{v}{\sqrt{1+u^2+v^2}}.
 \end{equation}

Hence, we have the equivalent flow 

\begin{widetext}
\begin{align}
& \frac{d U}{d\eta}=\frac{U V \sqrt{1-U^2-V^2}  \left(-2 b_{0}^2-b_{0} (2 \beta +1) e^{\frac{4 U}{\sqrt{1-U^2-V^2}}}+\beta  e^{\frac{8 U}{\sqrt{1-U^2-V^2}}}\right) \exp \left(-b_{0} e^{-\frac{4 U}{\sqrt{1-U^2-V^2}}}-\frac{2 U}{\sqrt{1-U^2-V^2}}\right)}{6 \left(b_{0}+\beta  e^{\frac{4
   U}{\sqrt{1-U^2-V^2}}}\right)^2} \nonumber \\
   & +\frac{1}{6} \rho_{0} U V (3 w_m+1) \sqrt{1-U^2-V^2}
   e^{-\frac{U (3 w_m+1)}{\sqrt{1-U^2-V^2}}}+V(1-U^2-V^2)+V^3, \label{syst:5.13}
  \\
  &  \frac{d V}{d\eta}=\frac{\left(V^2-1\right) \sqrt{1-U^2-V^2} \left(-2 b_{0}^2-b_{0} (2 \beta +1) e^{\frac{4 U}{\sqrt{1-U^2-V^2}}}+\beta  e^{\frac{8
   U}{\sqrt{1-U^2-V^2}}}\right)  \exp \left(-b_{0} e^{-\frac{4 U}{\sqrt{1-U^2-V^2}}}-\frac{2
   U}{\sqrt{1-U^2-V^2}}\right)}{6 \left(b_{0}+\beta  e^{\frac{4 U}{\sqrt{1-U^2-V^2}}}\right)^2}\nonumber \\
    & +\frac{1}{6} \rho_{0} e^{-\frac{U (3 w_m+1)}{\sqrt{1-U^2-V^2}}} \left(V^2-1\right) (3 w_m+1) \sqrt{1-U^2-V^2}-U V^2 \label{syst:5.14}.
\end{align}
For the special choice of parameters
$\frac{e^{-b_{0}} \left(2 b_{0}^2+2 b_{0} \beta +b_{0}-\beta \right)}{6 (b_{0}+\beta )^2}-\frac{1}{6} \rho_{0} (3 w_m+1)=0$,
  the origin is a fixed point of the dynamical system.
  Given
$
\gamma:= U_{\text{eff}}''(0)=\frac{e^{-b_{0}} \left(b_{0}^2 \left(-\left(8 b_{0}^2+2 b _{0}+(6 b _{0}+3) w_m+3\right)\right)+\beta ^2 (2
   (7-4 b _{0}) b _{0}+(3-6 b _{0}) w_m-1)-4 b _{0} \beta  (b _{0} (4 b _{0}+3
   w_m-3)-3)\right)}{6 (b _{0}+\beta )^3}$.
      \end{widetext}
      
      Depending of whether sign it has, the origin is a saddle ($U_{\text{eff}}''(0)$<0) or a focus ($U_{\text{eff}}''(0)>0$).

Assuming $\gamma>0$, and 
by taking a linear expansion around the origin, we find the approximate system
\begin{align}
& \frac{d U}{d\eta}=V, \quad \frac{d V}{d\eta}=-\gamma U,
\end{align}
  with solution  
   \begin{align}
   & U(\eta )= U_0 \cos \left(\sqrt{\gamma } \eta \right)+\frac{V_0 \sin \left(\sqrt{\gamma } \eta \right)}{\sqrt{\gamma }},\\
   & V(\eta )= V_0 \cos
   \left(\sqrt{\gamma } \eta \right)-\sqrt{\gamma } U_0 \sin \left(\sqrt{\gamma } \eta \right).
   \end{align}
     Finally we have the parametric solution
	 \begin{align}
 a=\exp \left(\frac{U}{\sqrt{1-U^2-V^2}}\right),
  \dot a= \frac{B_0 V}{\sqrt{1-U^2-V^2}},
  \end{align} 
  which is a periodic solution for $\gamma>0$.

     For $\gamma<0$, the solution is
      \begin{align}
   & U(\eta )= U_0 \cosh \left(\sqrt{-\gamma } \eta \right)+\frac{V_0 \sinh \left(\sqrt{-\gamma } \eta \right)}{\sqrt{-\gamma }},\\
   & V(\eta )= V_0 \cosh
   \left(\sqrt{-\gamma } \eta \right)-\sqrt{-\gamma } U_0 \sinh \left(\sqrt{-\gamma } \eta \right),
   \end{align}
  and has we commented before the origin is a saddle. 
  
On the other hand, in the same way  as for \eqref{Eq-4.15}-\eqref{Eq-4.16}, the asymptotic system near the fixed point at infinity for $U>0, w_m>0$ is 

 \begin{align}
 & \frac{d U}{d\eta}=V^3, \quad \frac{d V}{d\eta}=-U  V^2.
 \end{align}

 The orbits lying on the circle at infinity can be parametrized as
 $$U(\eta)= \pm \frac{c_2+\eta }{\sqrt{\left(c_2+\eta\right){}^2+1}}, \quad V(\eta)= \sqrt{\frac{1}{\left(c_2+\eta \right){}^2+1}}$$
    or 
    $$U(\eta)=\pm \frac{\eta -c_2}{\sqrt{\left(\eta -c_2\right){}^2+1}}, \quad V(\eta )= -\sqrt{\frac{1}{\left(\eta -c_2\right){}^2+1}}.$$
  Thus, there are fixed points lying on the circumference at infinity  $(U,V)=(\pm 1,0)$. 
  
\begin{figure*}[!ht]
\centering
\includegraphics[width=.6\textwidth]{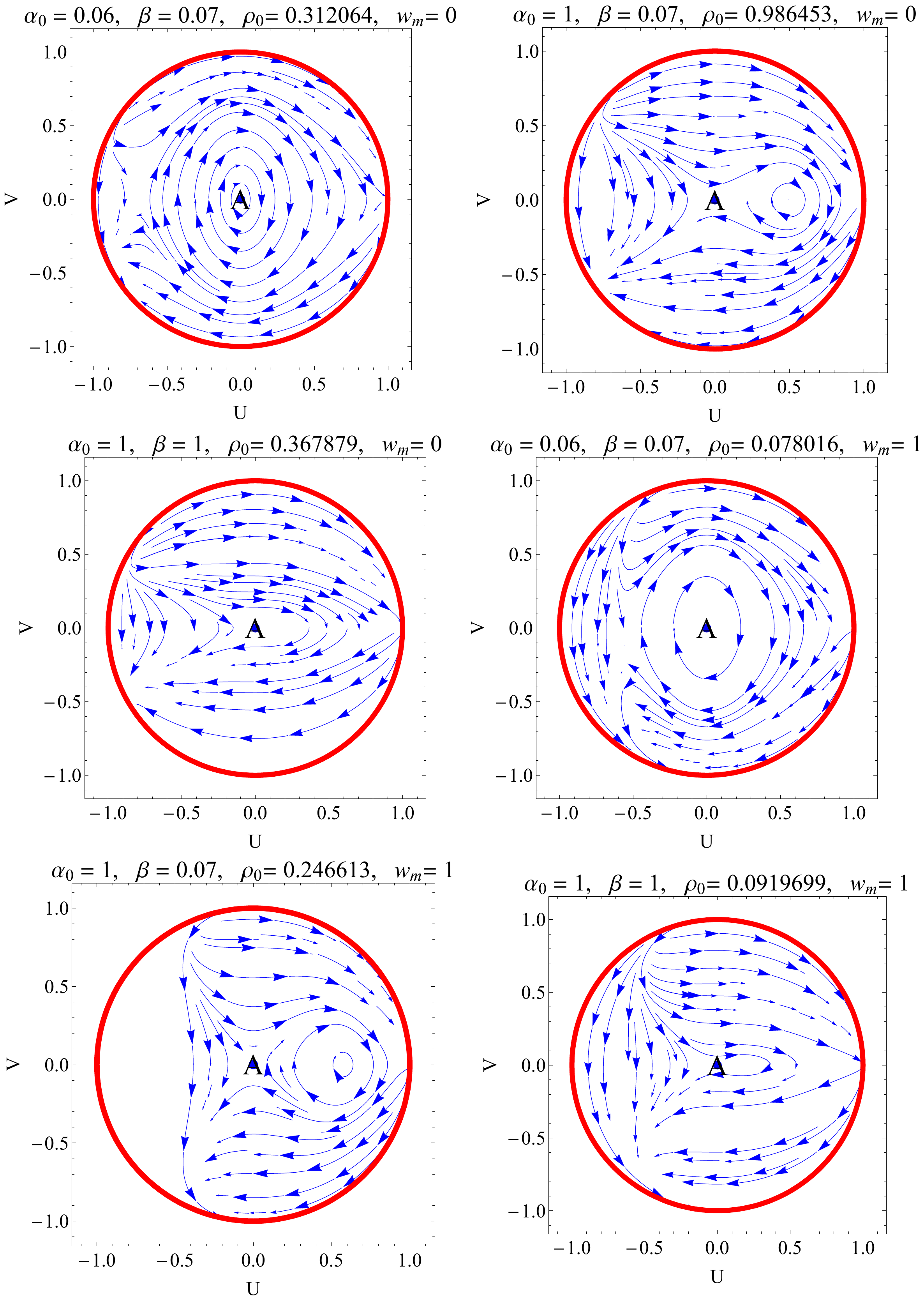}
\caption{\label{fig8} Dynamics of system  \eqref{syst:5.13}-\eqref{syst:5.14} for the  for some values of $b_0$ and $\beta$, and for a) dust $w_m=0$ and b) for stiff matter $w_m=1$. We have assumed $\frac{e^{-b_{0}} \left(2 b_{0}^2+2 b_{0} \beta +b_{0}-\beta \right)}{6 (b_{0}+\beta )^2}-\frac{1}{6} \rho_{0} (3 w_m+1)=0$.}
\end{figure*}

   Now, we proceed to the numerical integration of system \eqref{syst:5.13}-\eqref{syst:5.14}  for some values of $b_0$ and $\beta$, and for a) dust  and b) for stiff matter. There is a numerical evidence, as showed in in Fig. \ref{fig8}, of  the realization of a cyclic Universe in our model supported by NED. That is, the scale factor goes below and above the value $a=1$, reaching a maximum and a minimum value of $a$, and $a$ is bounded away zero, as expected from our NED proposal.  As shown in figure \ref{fig8} (b) and (e), when $A$ is a saddle, there are other cyclic solutions.

     \subsection{Integrability and connection with the observables}

In this section we comment on the integrability of the system at hand, and calculate some observables in terms of redshift. 
We assume that $B$ is not a constant or trivial.

By introducing the logarithmic time variable $N=\ln a$,  it is easy to recast the field equations as
\small{\begin{align}
& \frac{d B}{d N}= -2 B,\\
& \frac{d H}{d N}=\frac{B^2 e^{-\frac{1}{2} \alpha  B^2} \left(-4 \beta +\alpha ^2 B^4+2 \alpha  \beta  B^2\right)}{3 H \left(2 \beta +\alpha  B^2\right)^2}-\frac{1}{2} (w_m+1) \frac{\rho_{m}}{H},\\
& \frac{d\rho_{m}}{d N}= -3 (w_m+1) \rho_{m},\\
&
\frac{B^2 e^{-\frac{1}{2} \alpha  B^2}}{2 \beta +\alpha 
   B^2}+\rho_m=3 H^2.\label{eq_96}
   \end{align}}
As before $a=1$ corresponds to present value of the scale factor, so that $N=0$ today and $H(0)=H_0$, $B(0)=B_0$ and $\rho_{m}(0)=3 H_0^2 \Omega_{m0}$ are the current values of the Hubble factor, the magnetic field, and the matter density, respectively, where we have defined the current normalized energy density $\Omega_m(0)=:\frac{\rho_m}{3 H^2}|_{\text{today}}=\Omega_{m0}$. We use the dimensionless parameters $h_0=\alpha H_0^2, \quad b_0=\frac{\alpha B_0^2}{2}$.  

Integrating the above system and evaluating $N=\ln a$ and $a=\frac{1}{1+z}$, we obtain
\begin{widetext}
\begin{align}
& B(z)=B_0 (z+1)^2,\\
& \rho_m(z)=3 {H_0}^2\Omega_{m 0} (1+z)^{3(1+w_m)},
\\
& H(z)=H_0\sqrt{\frac{(1-\Omega_{m 0}) (z+1)^4 e^{-b_0 z (z+2) (z (z+2)+2)}}{3 h_0 (1-\Omega_{m 0}) z (z+2) (z (z+2)+2) e^{b_0}+1}+\Omega_{m 0} (z+1)^{3 (w_m+1)}},\label{eq:Hzwithom}
\\
&q+1=
\frac{\frac{4 b_0 (\Omega_{m0}-1)(z+1)^8 e^{b_0 
   \left(1-(z+1)^4\right)}}{{3 h_0 (\Omega_{m0}-1) \left(1-(z+1)^4\right) e^{b_0}}+1}-\frac{4 (\Omega_{m0}-1)e^{b_0 \left(1- (z+1)^4\right)} \left(3 h_0(\Omega_{m0}-1) e^{b_0}+1\right)}{\left(3 h_0 (\Omega_{m0}-1) \left(1-(z+1)^4\right)
   e^{b_0}+1\right)^2}+3 (w_m+1) \Omega_{m0} \left({z+1}\right)^{3( w_m+1)}}{2  \left(\Omega_{m0}
   \left({z+1}\right)^{3 (w_m+1)}-\frac{(\Omega_{m0}-1)(z+1)^4 e^{b_0 \left(1-(z+1)^4 \right)}}{3 h_0
   (\Omega_{m0}-1) \left(1-(z+1)^4 \right) e^{b_0}+1}\right)}.\label{eq112}
\end{align}

  \end{widetext}
Summarizing, the function $H(z)$ has the free parameters $H_0, \Omega_{m0}, h_0, b_0$ than can be contrasted against data and experiments. Then, we calculate the model parameters as:
\begin{equation}
\label{eq_108}
\alpha=\frac{h_0}{H_0^2},\; B_0=H_0 \sqrt{\frac{2 b_0}{h_0}},\; \beta=  b_0 \left(\frac{e^{-b_0}}{3h_0 (1-\Omega_{m0})}-1\right),
\end{equation}
and the current value of $q$ is
   \begin{align}
  & q_0= 1+\frac{1}{2} (3 w_m-1) \Omega_{m0} \nonumber \\
   & -2 (1-\Omega_{m0}) \left( b_0+3 h_0 (1-\Omega _{m0}) e^{b_0}\right).
   \end{align}

\section{Observational constraints}
\label{Sec. 6}

\begin{figure*}[!ht]
\centering
\includegraphics[width=0.6\textwidth]{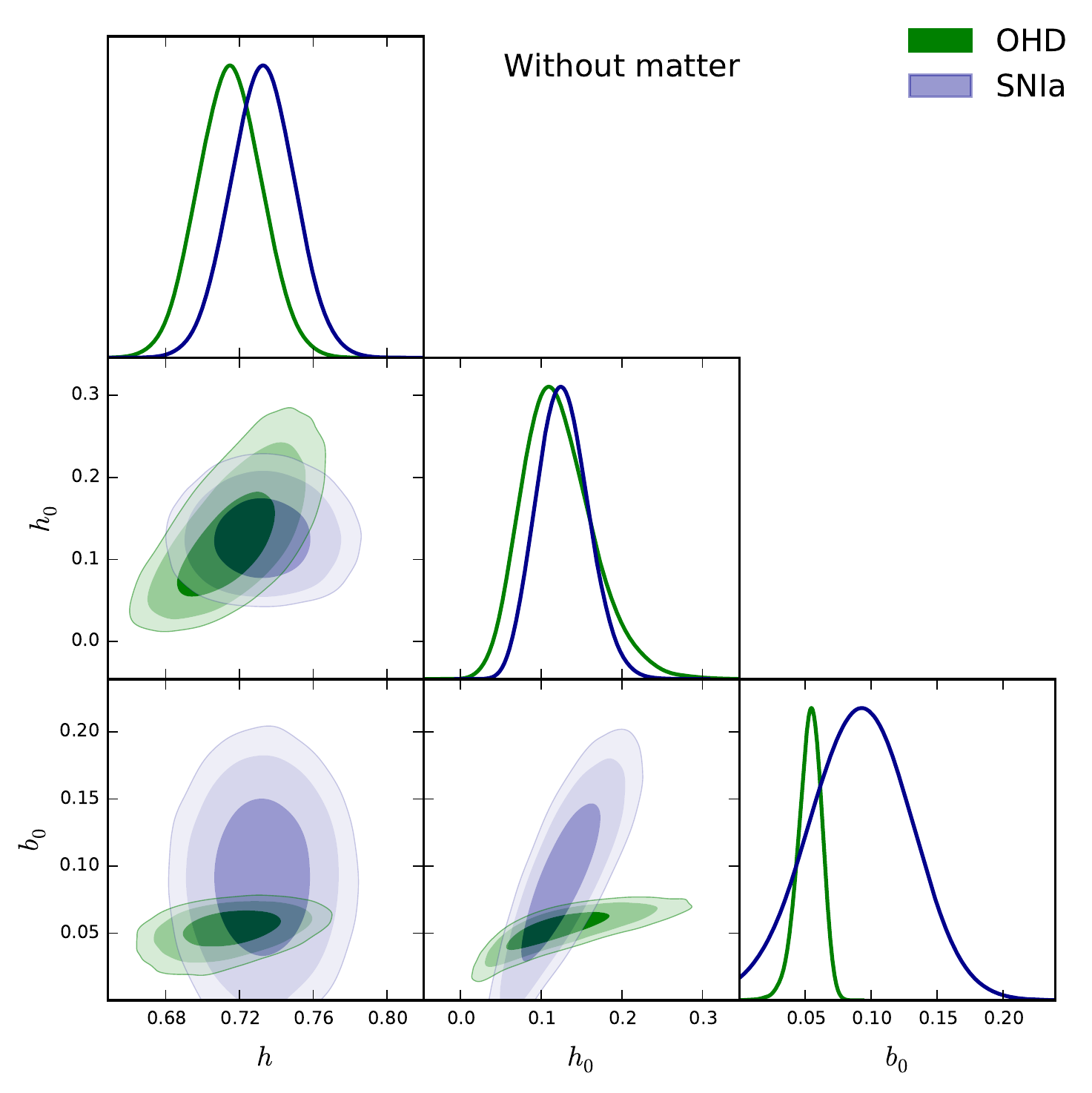}
\caption{1D marginalized posterior distributions and the 2D $68\%$, $95\%$, $99.7\%$ confidence levels 
for the $h$, $h_{0}$, and $b_{0}$ parameters of the NED cosmology without a matter field obtained from OHD and SNIa data.}
\label{fig:ned_contours_nomatter}
\end{figure*}

\begin{table*}[!ht]
\centering
\begin{tabular}{|ccccccccc|}
\multicolumn{9}{c}{NED model}\\
\hline
Data set & $\chi^{2}_{min}$ & $\Omega_{m0}$ & $h$ &$h_{0}$ &$b_{0}$&$B_{0} (10^{-31}\mathrm{cm^{-1}})$&$\alpha (10^{59} \mathrm{cm})$&$\beta$\\
\hline
\multicolumn{9}{|c|}{Without matter}\\
\multicolumn{9}{|c|}{}\\
OHD & $27.14$ & $---$ & $0.71^{+0.01}_{-0.01}$ & $0.11^{+0.04}_{-0.03}$& $0.05^{+0.008}_{-0.009}$ & $7.35^{+0.91}_{-0.73}$& $1.97^{+0.71}_{-0.61}$ &$0.08^{+0.05}_{-0.03}$\\
SNIa & $683.20$ & $---$ & $0.73^{+0.01}_{-0.01}$ & $9.39^{+1.29}_{-1.11}$& $0.60^{+0.40}_{-0.41}$ & $9.51^{+1.12}_{-1.47}$ & $2.01^{+0.50}_{-0.53}$ & $0.11^{+0.03}_{-0.03}$\\
\hline
\multicolumn{9}{|c|}{With matter}\\
\multicolumn{9}{|c|}{}\\
OHD & $15.10$ & $0.31^{+0.008}_{-0.008}$ & $0.73^{+0.01}_{-0.01}$ & $0.53^{+0.23}_{-0.24}$& $0.21^{+0.30}_{-0.15}$ & $7.05^{+7.92}_{-3.69}$& 
$8.44^{+3.66}_{-3.95}$ & $-0.02^{+0.05}_{-0.06}$\\
SNIa & $682.62$ & $0.31^{+0.009}_{-0.009}$ & $0.73^{+0.01}_{-0.01}$ & $0.37^{+0.09}_{-0.09}$& $0.11^{+0.11}_{-0.07}$ & $6.12^{+3.82}_{-2.89}$&$6.00^{+1.53}_{-1.59}$&$0.01^{+0.07}_{-0.01}$\\ 
\hline
\end{tabular}
\caption{Mean values for the NED parameters ($\Omega_{m0}$, $h$, $\alpha$, and $B_{0}$) 
derived from OHD and SNIa data (JLA sample). When the SNIa data are used, we estimate 
$M^{1}_{B}=-18.88^{+0.05}_{-0.05}\;(-18.94^{+0.05}_{-0.05})$, $\delta_{M}=-0.06^{+0.02}_{-0.01}$, $a=0.14^{+0.006}_{-0.006}$, and $b=3.10^{+0.08}_{-0.08}$ for no matter (matter) case.}
\label{tab:nedbf}
\end{table*}

In this section we test two NED models: without a fluid component and including it as dust matter (associated to a dark matter component), i.e. its equation of state is $w_{m}=0$. The free parameters $H_0, h_0, b_0$, and $\Omega_{m0}$ (when dust matter is considered) are constrained using the observational Hubble data from cosmic chronometers and the latest SNIa data. Next we calculate the model parameters using Eqs. \eqref{eq_74} and \eqref{eq_108}.

\begin{figure*}
\centering
\includegraphics[width=0.6\textwidth]{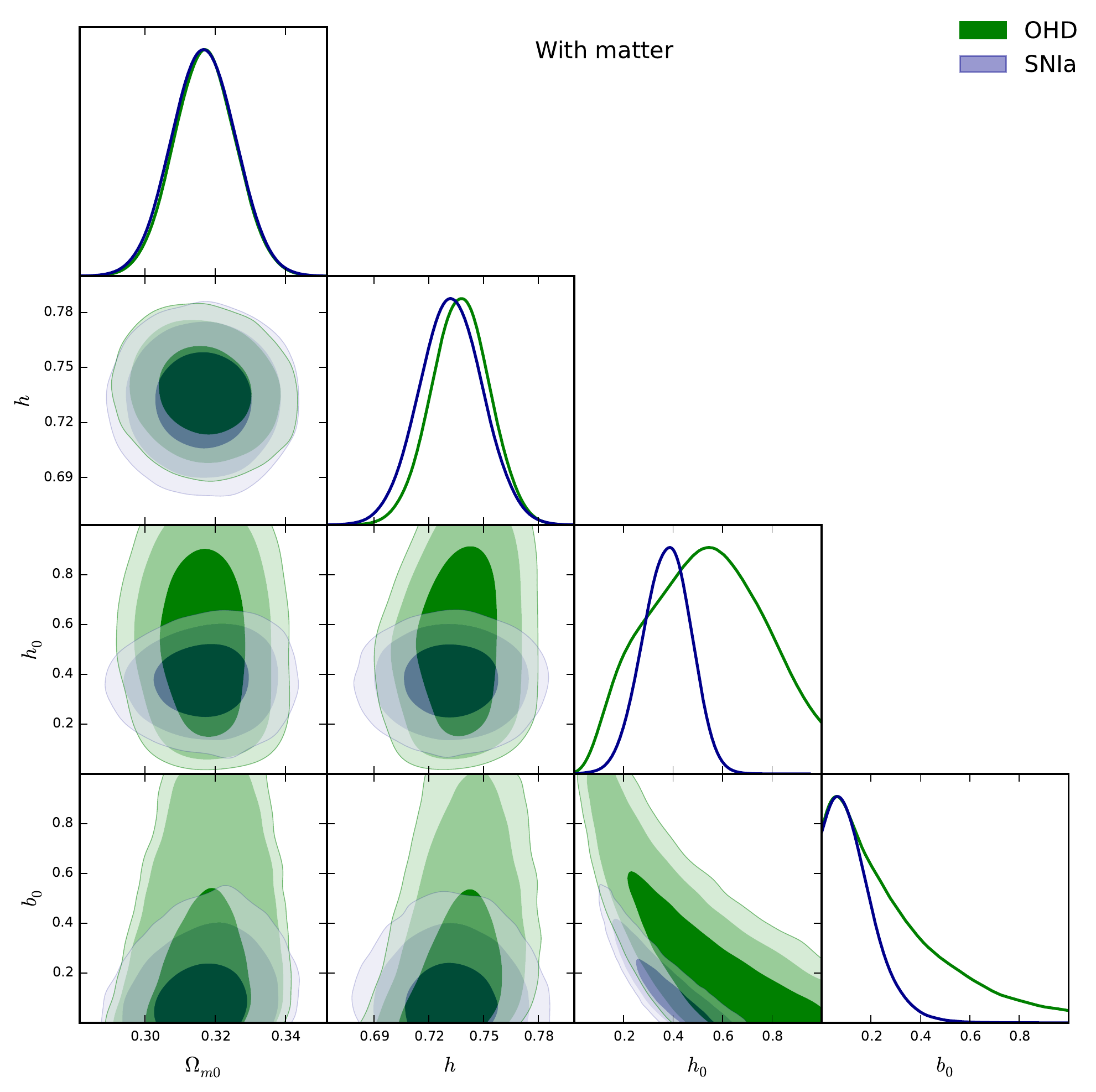}
\caption{1D marginalized posterior distributions and the 2D $68\%$, $95\%$, $99.7\%$ confidence levels 
for the $\Omega_{m0}$, $h$, $h_{0}$, and $b_{0}$ parameters of the NED cosmology including dust matter obtained from OHD and SNIa data.}
\label{fig:ned_contours_matter}
\end{figure*}

\begin{figure}[ht!]
\begin{center}
\includegraphics[scale=0.4]{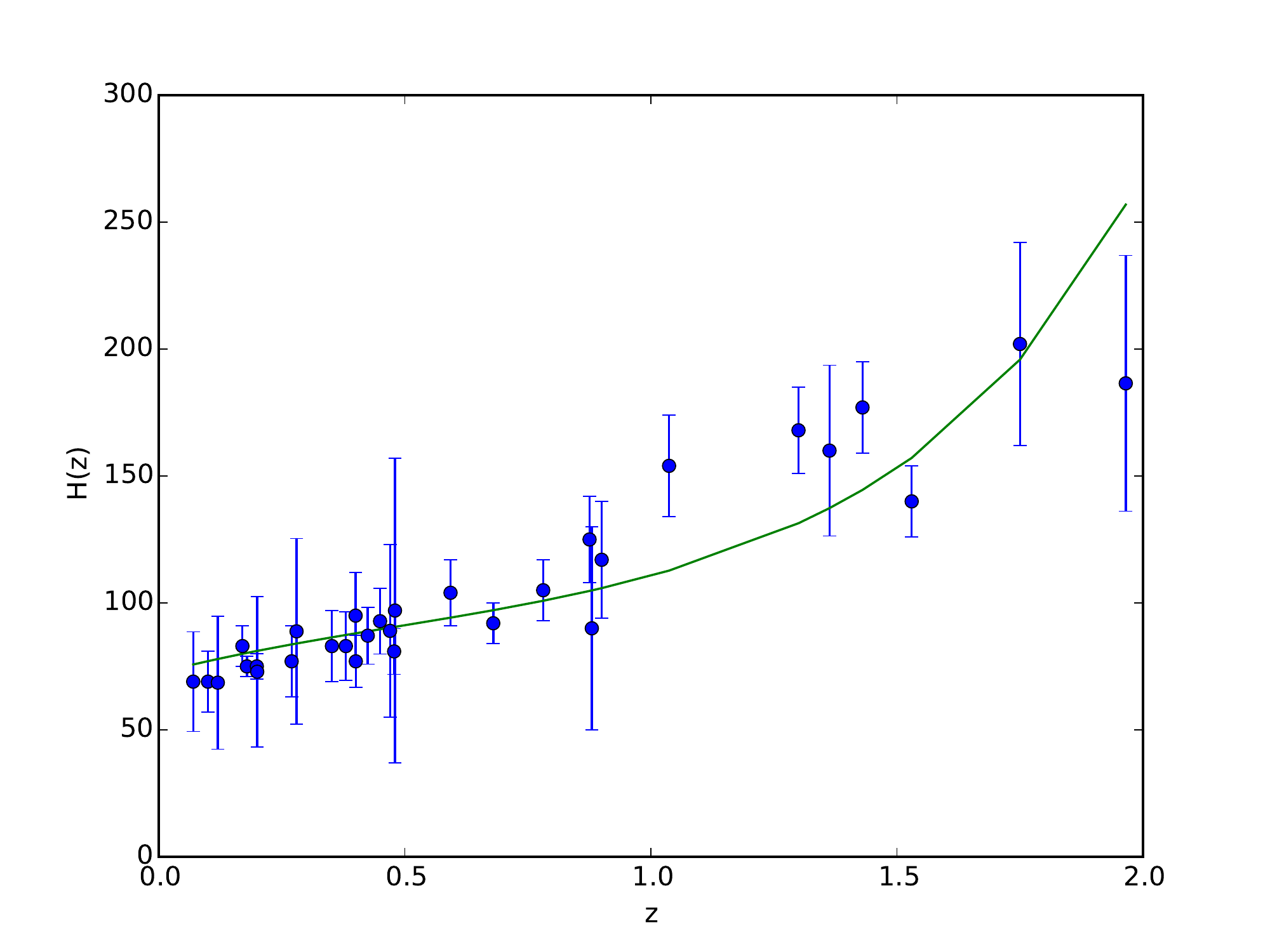}
\includegraphics[scale=0.4]{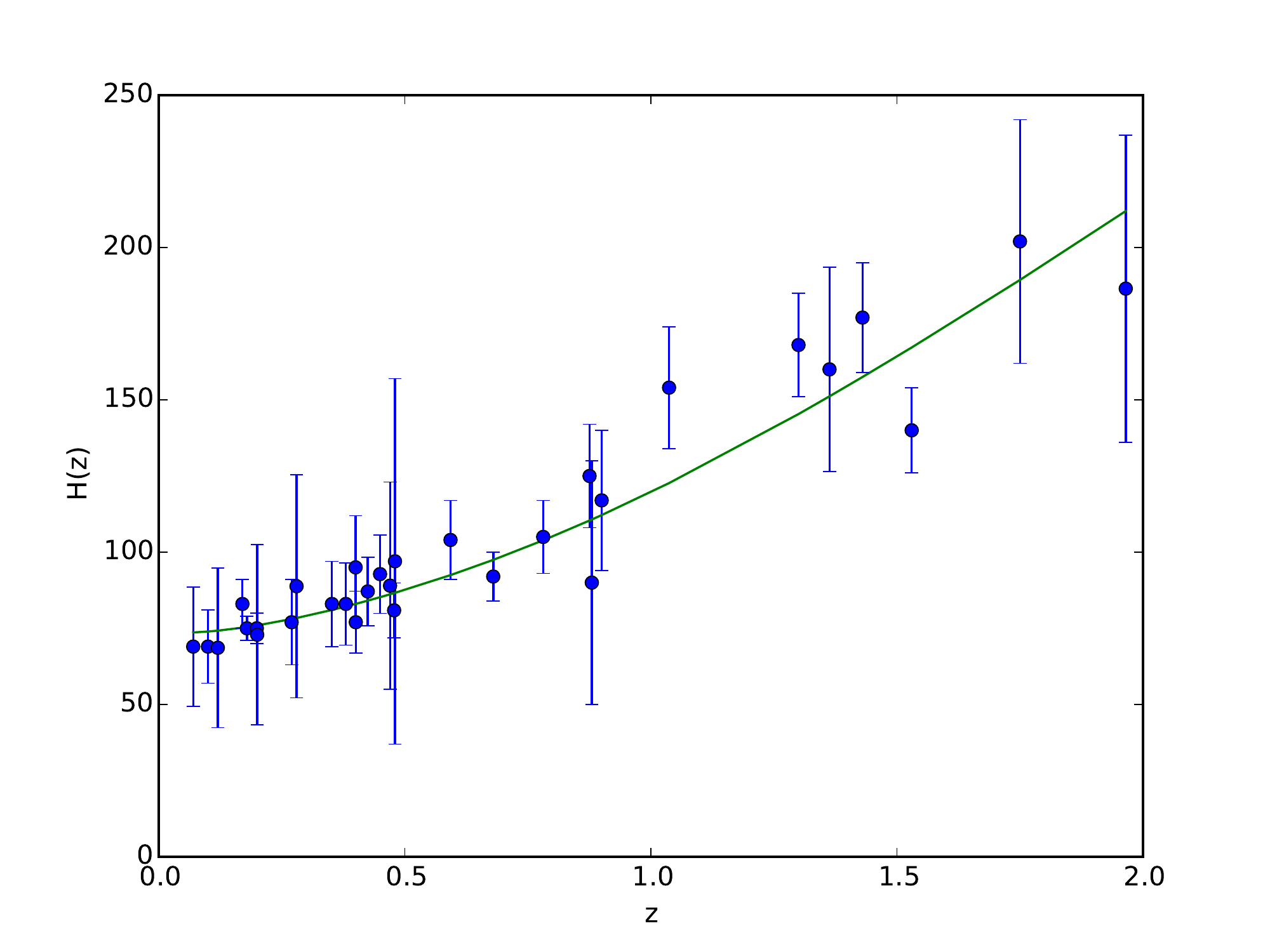}
\caption{Fitting to OHD data (blue dots) using the mean values of the NED parameters into the theoretical $H(z)$ (solid line) for the case without matter (Eq. \ref{eq:Hznomatter}, top panel) and including matter (Eq. \ref{eq:Hzwithom}, bottom panel)}
\label{fig:hz_fit}\end{center}
\end{figure}
\begin{figure}[ht!]
\begin{center}
\includegraphics[scale=0.35]{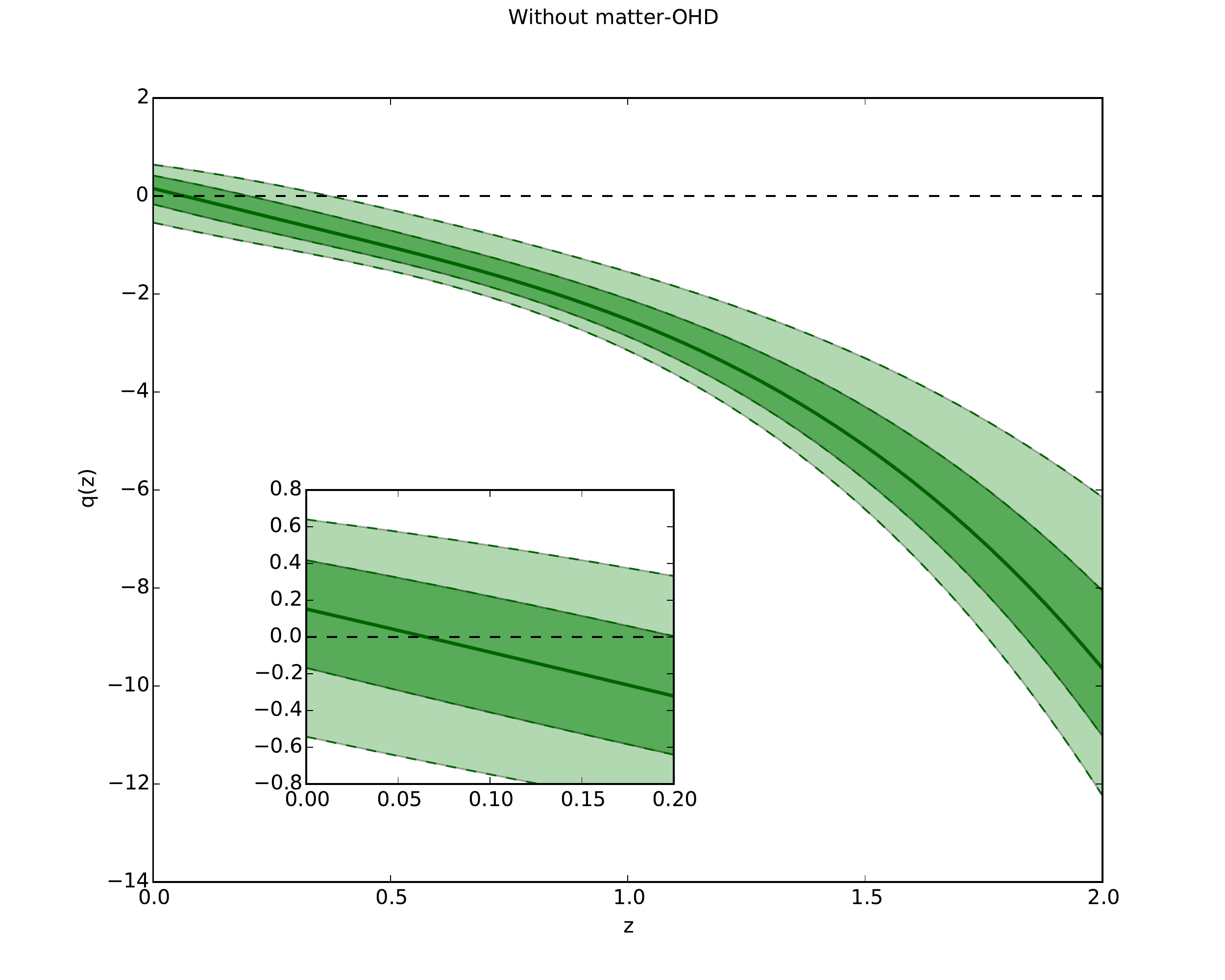}
\includegraphics[scale=0.35]{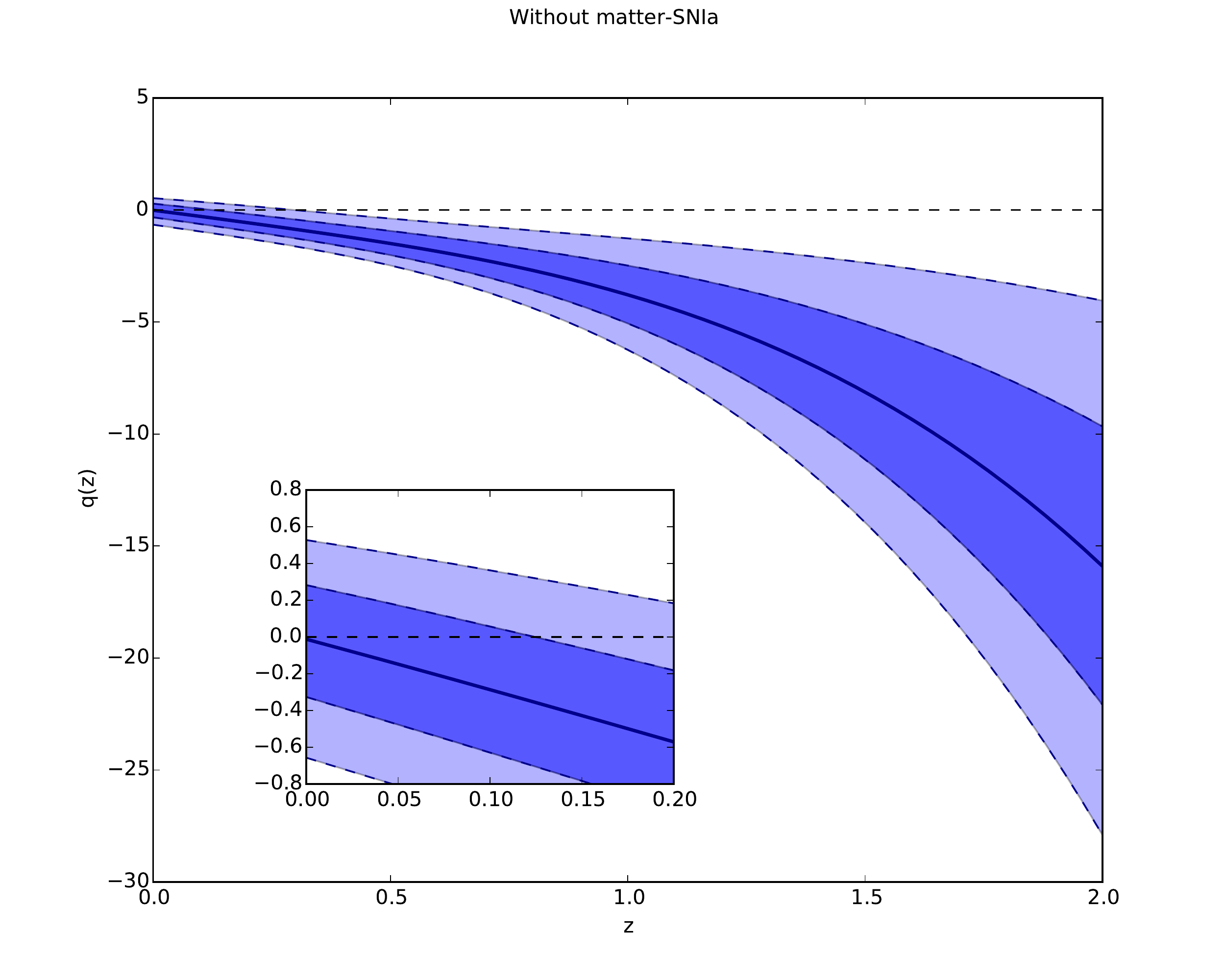}
\caption{The top and bottom panels show the reconstruction of the deceleration parameter $q(z)$ using the $h$, $h_{0}$, and $b_{0}$ mean values from OHD and SNIa data respectively into the theoretical $q(z)$ given by \eqref{eq:qznomatter}. The shadowed areas represent the $1\sigma$ and $2\sigma$ confidence levels. The inset shows the $q(z)$ behavior in the range $0<z<0.2$.}
\label{fig:qznomatter}\end{center}
\end{figure}
\begin{figure}[ht!]
\begin{center}
\includegraphics[width=9cm,scale=0.45]{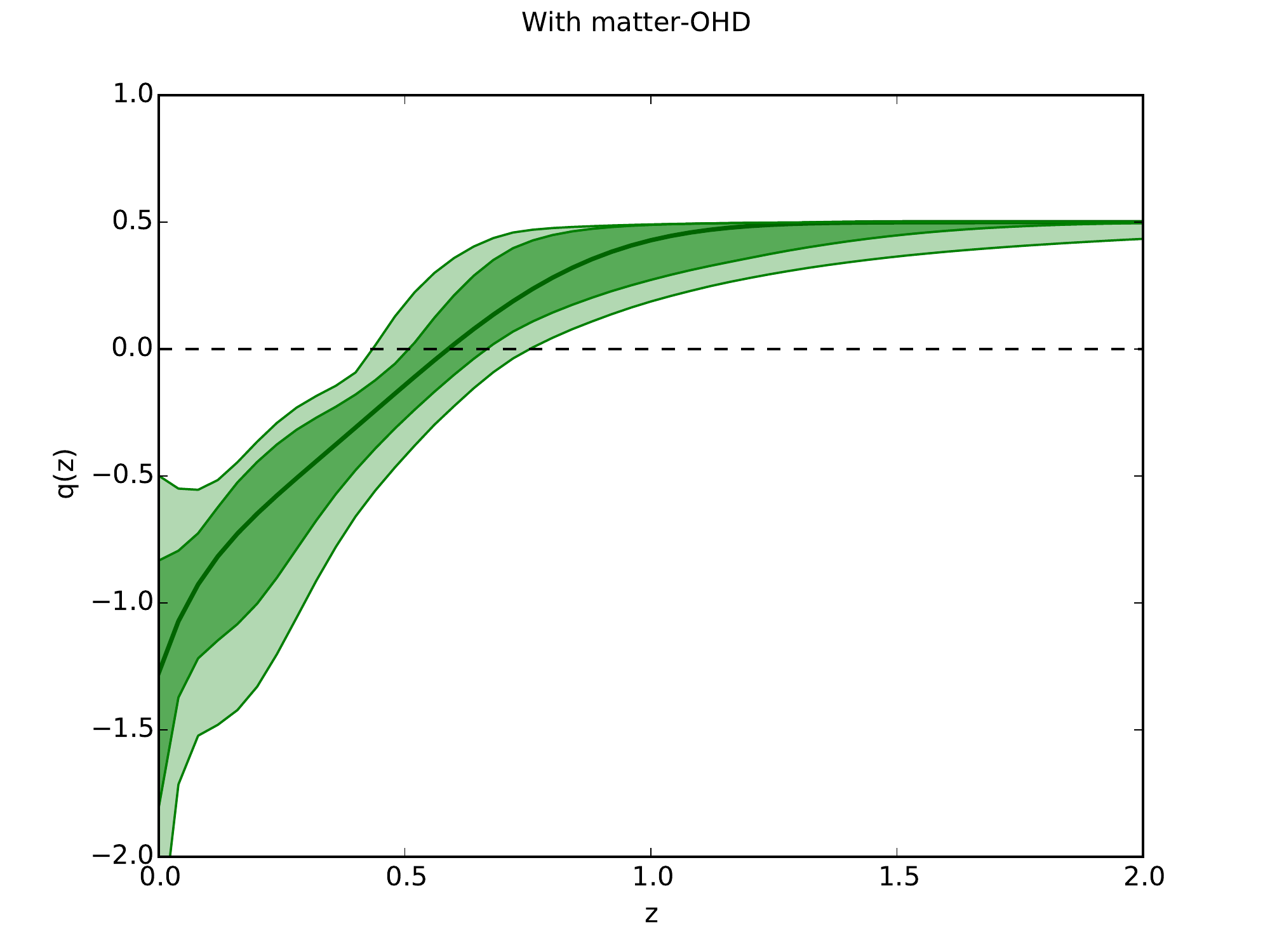}
\includegraphics[width=9cm,scale=0.45]{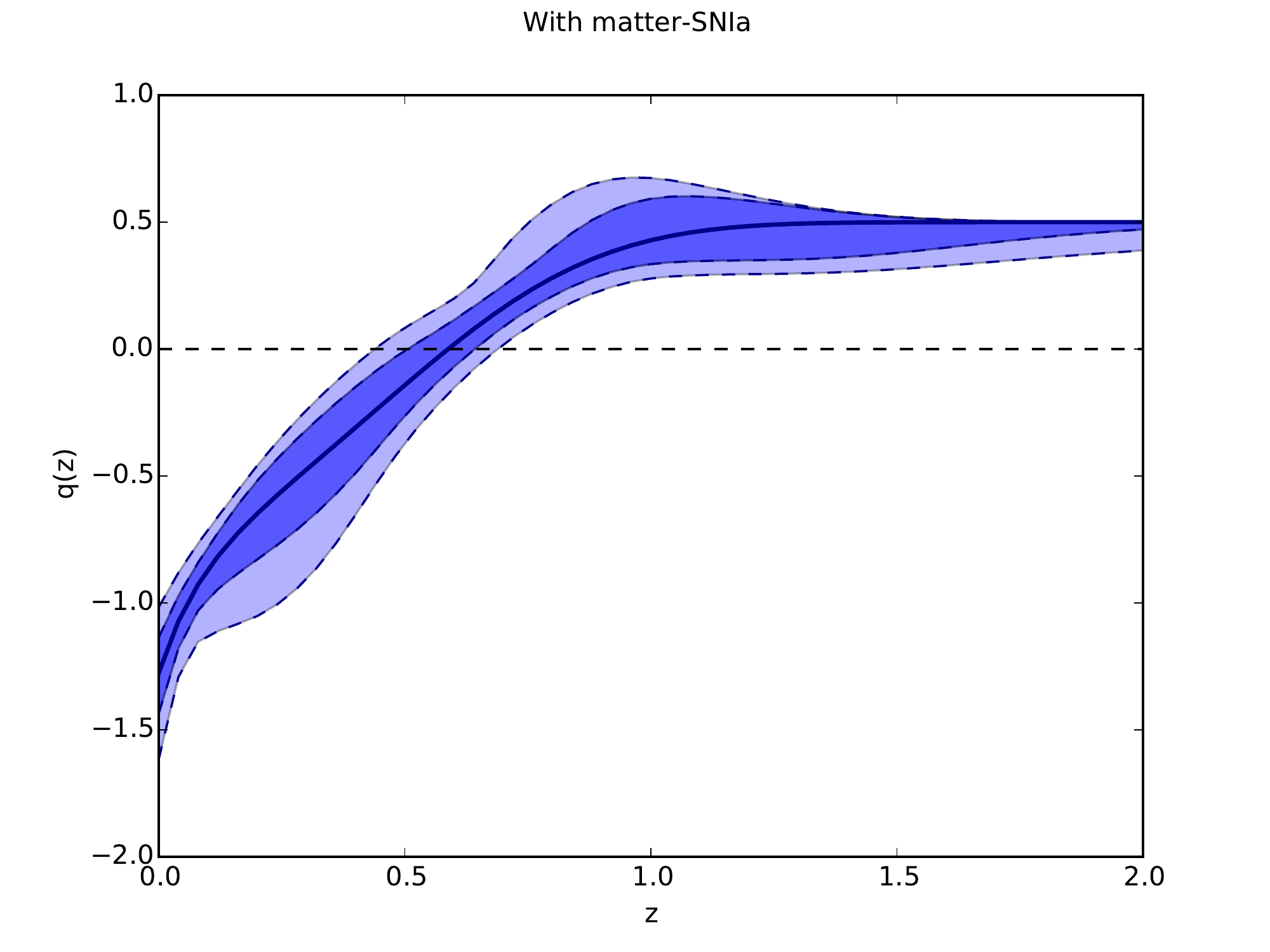}
\caption{The top and bottom panels show the reconstruction of the deceleration parameter $q(z)$ using using the $\Omega_{m0}$, $h$, $h_{0}$, and $b_{0}$ mean values from OHD and SNIa data respectively into the theoretical $q(z)$ given by \eqref{eq112}. The shadowed areas represent the $1\sigma$ and $2\sigma$ confidence levels.}
\label{fig:qzmatter}
\end{center}
\end{figure}

\begin{itemize}
\item{Observational Hubble data (OHD)}.
The differential age (DA) method measures $H(z)$ between two passively-evolving galaxies with similar metallicities and separated by a small redshift interval (cosmic chronometers) \cite{Jimenez:2001gg,Moresco:2012}. The data provided by the DA method are cosmological-model-independent and then they can be used to probe alternative cosmological models. Here, we use 
the latest OHD obtained from DA technique, which contains $31$ data points covering $0 < z < 1.97$, compiled by \cite{Magana:2017} and references therein.
The figure-of-merit for the OHD is written as
\begin{equation}
\chi_{\mbox{OHD}}^2 = \sum_{i=1}^{31} \frac{ \left[ H(z_{i}) -H_{da}(z_{i})\right]^2 }{\sigma_{H_i}^{2}}
+\left(\frac{H_{0}-73.24}{1.74}\right)^{2},
\end{equation}
where $H(z_{i})$ is the theoretical Hubble parameter (given by either Eq. \ref{eq:Hznomatter} or Eq. \ref{eq:Hzwithom}, depending on whether we consider no matter or we include it, respectively),
$H_{da}(z_{i})$ is the observational one at redshift $z_{i}$, and
$\sigma_{H_i}$ its error. Notice that in the last expression, we also consider
the measurement of $H_{0}=73.24\pm1.74\mathrm{Km s^{-1}Mpc^{-1}}$ by \cite{Riess:2016jrr} as a Gaussian prior.

\item{Type Ia Supernovae.} The first evidence of the accelerating expansion of the Universe was provided by
the observations of distant type Ia Supernovae (SNIa) \cite{riess,perl}. Over the last years, the high-resolution
SNIa observations have demonstrated be a key cosmological probe due the shape their light curves can be standardizable. Thus, any alternative cosmological model should be confronted with the latest SNIa data.
Here, we use the joint-light-analysis (JLA) compilation by \cite{Betoule:2014} consisting in $740$ data points
in the range $0.01<z<1.2$. For the JLA sample, the observational distance modulus can be computed as
\begin{equation}
\mu_{obs}=m_{B}-\left(M_{B}-a\,X_{1}+b\,{C}\right),
\label{eq:muobs}
\end{equation}
where $m_{B}$ is the observed peak magnitude in rest-frame B band, $X_{1}$ is the time stretching of the light-curve,
$C$ is the supernova color at maximum brightness, and $M_{B}$\footnote{Notice that\begin{equation}
    M_{b}=
    \begin{cases}
      M^{1}_{b}, & \text{if the host stellar mass} \,M_{stellar}<10^{10}M_{\odot} \\
      M^{1}_{b}+\delta_{M}, & \text{otherwise}
    \end{cases}
  \end{equation}}, $a$, and $b$ are nuisance parameters in the
distance estimate. On the other hand, the theoretical distance modulus is given by $\mu_{th}=5\log_{10}(d_{L}/10pc)$,
where $d_{L}$, the luminosity distance predicted by the NED cosmology (an interesting 
study of the SNIa luminosity distance in Born-Infeld NED cosmology is presented in \cite{Aiello:2008}), reads as
\begin{equation}
d_{L}=(1+z)c \int^{z}_{0}\frac{\mathrm{dz}^{\prime}}{H(z^{\prime})}.
\label{eq:dl}
\end{equation}

Therefore, the figure-of-merit for the SNIa data can be written as
\begin{equation}
\chi^{2}_{\mbox{SNIa}}=\mathbf{\left(\mu_{obs} - \mu_{th}\right)^{\dag}}\mathrm{Cov}(a,b)^{-1}\mathbf{\left(\mu_{obs} - \mu_{th} \right)},
\end{equation}
where $\mathrm{Cov}(a,b)$ is the covariance matrix \footnote{available  at  \url{http://supernovae.in2p3.fr/sdss_snls_jla/ReadMe.html}} of $\mathbf{\mu_{obs}}$ provided by  \cite{Betoule:2014}.
\end{itemize}

To constrain the  free parameters, ($\Omega_{m0}$, $h$, $h_0$, and $b_{0}$), we perform a Bayesian Monte Carlo Markov Chain (MCMC) analysis employing
the emcee python module using $800$ walkers, $500$ steps in the burn-in-phase, and $4000$ MCMC steps to guaranty the convergence.  We consider a Gaussian prior on $\Omega_{m0}$ as measured by \cite{Planck:parameters} and uniform priors $[0,1]$ for both $h_{0}$ and $b_{0}$.

Figures \ref{fig:ned_contours_nomatter} and \ref{fig:ned_contours_matter} show the $1$D marginalized posterior distribution and the 2D confidence contours for the $h$, $h_{0}$, and $b_{0}$ parameters for the NED cosmology without matter and $\Omega_{m0}$, $h$, $h_{0}$, and $b_{0}$ when the matter component is included respectively. Notice that both OHD and SNIa data provide consistent constraints on the NED parameters for both models.

Table \ref{tab:nedbf} gives the mean values
for the NED model parameters using different cosmological data for both cases: without and including a matter component. The chi-square values indicate a over-fitting to OHD but a good-fitting to SNIa data. All our constraints on the current magnetic field, $B_{0}\sim 10^{-31}\mathrm{cm^{-1}}$, are larger than the upper limit $10^{-33}\mathrm{cm^{-1}}$ by the Planck satellite implying that NED cosmologies could not be suitable to explain the Universe dynamics at late times. Nevertheless,
the Figure \ref{fig:hz_fit} illustrates a good fitting of the NED cosmology without (top panel) and with (bottom panel) matter to OHD. In addition, the Figures \ref{fig:qznomatter} and \ref{fig:qzmatter} show the reconstructed deceleration parameter $q(z)$ in the range $0<z<2$ for each data set without matter and including matter respectively. If the matter component is not included, although the data sets predict an accelerated phase in the early Universe, a non accelerated Universe is preferred in the current epoch. However, a late cosmic acceleration dynamics is allowed within the $2\sigma$ confidence levels. When a matter component is included in the NED cosmology, the data set predict a $q(z)$ dynamics similar to that of the standard model. Indeed, $q(z)\rightarrow1/2$ when $z\rightarrow \infty$.   Moreover, both cosmological data favor up to $2\sigma$ confidence levels an accelerating expansion in the current epoch, i.e., the Universe passes of a decelerated phase to an accelerated stage at redshift $\sim 0.6$ for the OHD (top panel) and SNIa (bottom panel) constraints. Therefore, although the NED cosmology including dust matter ($w_{m}=0$) predict a higher $B_{0}$ value, it is able to drive a late-time cosmic acceleration which is consistent with the $Y-\beta$ allowed regions of the Figure \ref{fig:yvsbeta}.

\section{Conclusion}
\label{Sec. 7}
In this paper, we have considered a new field of NED as a source of gravity to shed light on the dynamics behind the accelerating Universe and solve the singularity problem of the Big Bang and the Universe curvature. 

We have argued that the NED on cosmological scales could reveal the acceleration of the Universe during the inflationary era. We have also shown that, after this period of cosmic inflation the Universe undergoes decelerated expansion and asymptotically approaches the Minkowski spacetime.  

From the dynamical system approach we have found for parameters  $\beta= \frac{2 b_{0}^2+b_{0}}{1- 2 b_{0}}, b_{0}=0.06$, $B_0=0.5, U_0=V_0=0.2$, and assuming no background matter, the realization of a cyclic Universe in our model supported by NED near the value $a_0=1$. The scale factor goes below and above the value $a=1$, reaching a maximum and a minimum value of $a$, and $a$ is bounded away zero (there is no initial singularity, as expected from our NED proposal). For our two models (not including matter, and for prefect fluid),
We have found approximated solutions in parametric form for the scalar factor and its time derivative, which are valid in the neighborhood of the fixed point at the finite region. For some values of the parameters we found also the realization of a cyclic Universe. 

For the model without matter, we have found solutions with $B$ constant ($\alpha B^2= -\beta +\sqrt{\beta^2 + 4\beta}$), which supports the inflation similar to de-Sitter Universe, with 
$a\propto e^{\sqrt{\Lambda} t}, \Lambda=\frac{e^{\frac{1}{2} \left(\beta -\sqrt{\beta  (\beta +4)}\right)} \left(\beta
   -\sqrt{\beta  (\beta +4)}+2\right)}{6 \alpha }, \beta >0, \alpha >0$.

In the case of evolving $H$, it was provided explicit expressions for $H(z)$ by direct integration of the equations of motion in the two models: with and without including a perfect fluid. We have tested these NED models by performing a Bayesian Monte Carlo Markov Chain (MCMC) analysis using OHD and SNIa data.
We have found that the first case predicts $B_{0}$ values higher limits than the Planck satellites  bounds. That is $B_{0}\sim 10^{-31}\mathrm{cm^{-1}}$, are larger than the upper limit $10^{-33}\mathrm{cm^{-1}}$. In addition, although NED without matter is able to fit the OHD, it prefers no late cosmic acceleration. For the model including a fluid, we consider dust matter with an equation of state $w_{m}=0$ which is associated to a dark matter component. In this particular example we conclude that this is a good model for the early time Universe, but there are not statistical differences with the usual model for the radiation epoch. We have reconstructed the deceleration parameter $q(z)$ in the range $0<z<2$ as shown in Figure \ref{fig:qzmatter} for OHD (top panel) and SNIa data (bottom panel). Notice that $q(z)\rightarrow1/2$ when $z\rightarrow \infty$. In addition, both cosmological data predict an accelerating expansion, i.e., the Universe passes of a decelerated phase to an accelerated stage at redshift $\sim 0.6$. Nevertheless, the $B_{0}$ estimation is higher than the upper limit by Planck measurements. That is, first we have proved that a model based on NED alone does not passes the late cosmic acceleration test.
Second, we have proved that dust matter plus NED passed this test, and could provide late-time acceleration (the combined effect of both matter contributions).

Summarizing, using in a combined way the powerful of phase-space analysis, and the observational fit, one is able to falsify (in the sense of test) a theoretical cosmological model based on NED. From one side,  dynamical systems tools allows to identify regions in the parameter space to provide stability conditions for fixed points with physical meaning.  Furthermore, it was possible to use 2D phase spaces,  showing trajectories in a geometrical way, so that it becomes easy to observe the property with the help of the attractors which are the most easily seen experimentally. On the other hand, comparison with data allows to refine more the region of parameters. In principle, although the NED cosmology including dust matter predict a value of $B_{0}$ greater than the upper bound found by Planck satellite,  it is able to drive a late-time cosmic acceleration which is consistent with our dynamical systems analysis and it is preferred by OHD and SNIa data sets. In our opinion this is a very interesting result and deserves discussion.

\section*{Acknowledgements}
This work was supported by the Chilean FONDECYT Grant No. 3170035 (A\"{O}) and FONDECYT Grant No. 3160674 (JM). GL thanks
to Department of Mathematics at Universidad Catolica del Norte for warm hospitality and financial support. A. \"{O}. is grateful to Prof. Douglas Singleton for hosting him at the California State University, Fresno and also A. 
\"{O}. would like to thank Prof. Leonard Susskind and Stanford Institute for Theoretical Physics for hospitality.

\end{document}